\documentclass[article,prb,aps,twocolumn,amsmath,amssymb,floatfix,
superscriptaddress]{revtex4-2}
\usepackage[dvips]{graphics}
\usepackage[justification=raggedright]{caption}
\usepackage{subcaption}
\usepackage[font=footnotesize,figurewithin=none]{caption}
\usepackage{subcaption}
\usepackage{adjustbox}
\usepackage[labelfont=bf]{caption}
\usepackage[labelformat=parens, labelfont=bf]{subcaption}
\usepackage{color}
\usepackage{float}
\usepackage{natbib}
\usepackage[dvipsnames]{xcolor}
\setcitestyle{numbers}
\setcitestyle{square}
\definecolor{dred}{rgb}{0,0,0.6}
\usepackage{graphicx}    
\usepackage{bm}  
\usepackage{amssymb} 
\usepackage{mathtools}
\usepackage{mathdots}
\usepackage{amsmath,,amsfonts,bm}
\usepackage{braket}
\usepackage[mathscr]{euscript}
\usepackage[colorlinks=true, allcolors=blue]{hyperref}
\usepackage{url}
\begin{document}
\title{Topological characterization of a non-Hermitian ladder via Floquet non-Bloch theory
}

\author{Koustav Roy, Koustabh Gogoi, and Saurabh Basu \\ \textit{Department of Physics, Indian Institute of Technology Guwahati-Guwahati, 781039 Assam, India}}

\date{\today}
\begin{abstract}
In this paper, we study a non-Hermitian (NH) ladder subjected to a variety of driving protocols. The driven system looses chiral symmetry (CS) whose presence is indispensable for its topological characterization. Further, the bulk boundary correspondence (BBC) gets adversely affected due to the presence of non-Hermitian skin effect (NHSE). Here, we present a formalism that retrieves the lost CS, and subsequently restores the BBC via the construction of a generalized Brillouin zone (GBZ). Specifically, we employ delta and step drives to compare and contrast between them with regard to their impact on NHSE. Further, a widely studied harmonic drive is invoked in this context, not only for the sake of completeness, but its distinct computational framework offers valuable insights on the properties of out-of-equilibrium systems. While the delta and the harmonic drives exhibit unidirectional skin effect in the system, the step drive may show bi-directional skin effect. Also, there are specific points in the parameter space that are devoid of skin effect. These act as critical points that distinguish the skin modes to be localized at one boundary or the other. Moreover, for the computation of the non-Bloch invariants, we employ GBZ via a pair of symmetric time frames corresponding to the delta and the step drives, while a high-frequency expansion was carried out to deal with the harmonic drive. Finally, we present phase boundary diagrams that demarcate distinct NH phases obtained via tracking the trajectories of the exceptional points. These diagrams demonstrate a co-existence of the zero and $\pi$ energy modes in the strong NH limit and thus may be relevant for studies of Floquet time crystals.
\end{abstract}

\maketitle

\begin{center}\section{\label{sec:level1}Introduction}\end{center}

Since the discovery of the integer quantum Hall effect \cite{iqhe}, the study and control of different states of matter have become a key focus in condensed matter physics. The ability to prepare systems in topologically protected states has not only deepened our understanding of non-trivial phases but also opened up new possibilities for quantum information devices. One promising way to achieve such non-trivial (and even richer) properties is through the application of time-periodic driving. In this regard, Floquet engineering \cite{floquetformalism1,floquetformalism2,floquetformalism3} has emerged as a powerful quantum control technique, offering a compelling approach for fine-tuning topological properties in out-of-equilibrium systems. The formalism demonstrates that periodic driving can give rise to additional non-trivial states of matter that have no equivalence in static conditions. Furthermore, as a result of the time periodicity, the energy bands can be folded back into a Floquet Brillouin
zone (FBZ), at the boundary of which the system facilitates the emergence of anomalous topological states,
referred to as the $\pi$ modes \cite{floquet1,floquet2,floquet3,floquet4}. This concept has been
efficiently utilized in numerous experiments involving ultracold atoms in optical lattices, and photonic waveguides \cite{ultracoldfloquet1,ultracoldfloquet2,ultracoldfloquet3,ultracoldfloquet4,FloquetQSH,ultracoldfloquet6}. The flexibility of driving protocols in Floquet engineering opens up exciting new avenues for the discovery of exotic phenomena that are unattainable in static systems. This encompasses features such as, Floquet topological phase (FTP) with multiple edge states \cite{1dfloquet1,1dfloquet2,chern1,rashba,creutzfloquet,creutzfloquet2}, Floquet transport properties in planar junctions \cite{josephson1,josephson2,josephson3}, emergence of Floquet Anderson phase containing localized bulk with protected edge modes in quasiperiodic systems \cite{dimerizedkitaevfloquet}, emergence of discrete time crystalline phases \cite{period2t1,period2t2}, Floquet topological characterization of quantum chaos model \cite{dkr,khm} etc. These examples highlight the remarkable potential of Floquet engineering, revealing an expansive landscape of novel phenomena with seemingly endless possibilities.
\par There are variety of techniques for creating time periodicity and establishing topologically protected edge states, such as, illuminating matter with light \cite{shininglight1,shininglight2,shininglight3}, shaking optical lattices \cite{shakenoptical1,shakenoptical2}, and utilizing photonic systems \cite{photonicFloquet}. Among the studies of Floquet topological states, one-dimensional (1D) systems have been prominently featured due to their simplicity, such as the Su-Schrieffer–Heeger (SSH) model \cite{ssh} and its generalized variants \cite{genssh1,genssh2,genssh3} being notable examples. Some of these models hold promise as building blocks for topological fault-tolerant quantum computation \cite{nonabelian}. Periodic driving on such 1D systems has been shown to reveal a wealth of topological features across both the high \cite{creutzfloquet,lago,1dfloquet1} and the low-frequency regimes \cite{lowfreqfloquet}. In this context, and of relevance to us, the Creutz ladder \cite{creutz1,creutz2} is particularly noteworthy due to its practical realization in cold atomic systems \cite{creutzexpt1,creutzexpt2}. Interestingly, ladder networks belonging to Creutz family are valuable for uncovering significant two-dimensional topological aspects and symmetry classification schemes, owing to their quasi-1D nature \cite{hugelchiral}.  The model features two rungs of lattice sites connected by diagonal, vertical, and horizontal hoppings. Additionally, a magnetic flux threads the plane of the ladder, introducing an extra degree of freedom through the Peierls phase associated with the horizontal hopping. The localization of the zero-energy modes in this system is also influenced by Aharonov-Bohm caging. This dual protection arising from the Aharonov-Bohm caging and topological symmetry conservation ensures that the edge modes remain robust. Recent investigations propose various modified versions of the Creutz ladder, that hold rich information in the field of localization dynamics \cite{kuno,liberto}, many-body interactions \cite{juneman} etc. These compelling properties make the Creutz ladder a promising candidate for the exploration of fundamental physics and its applications in Floquet engineering.
\par In practical experimental setups, achieving such FTPs often requires the system to experience gain and loss \cite{gainandloss1,gainandloss2,gainandloss3} or environment-induced dissipation \cite{dissipation}. These processes can be effectively incorporated by introducing non-Hermiticity into the Hamiltonian of the Floquet system. Consequently, the dynamics governed by a time-periodic non-Hermitian (NH) Hamiltonian become non-unitary, and the quasienergies associated with the Floquet operator may become complex. The presence of complex energies and the non-orthogonality of eigenstates have led to significant advancements in the study non-Hermitian physics \cite{NHtheory1,NHtheory2,NHtheory3, dipendu1,dipendu2, NHtheory5}. A notable example of this phenomena is the emergence of exceptional points (EP) where a spectral degeneracy is accompanied by a coalescence of the corresponding eigenstates \cite{EP1,EP2,EP3,EP4,EP5}. Meanwhile, a substantial amount of research has been dedicated to verify the robustness of NH topological phases through experiments in fields, such as photonics \cite{NHphotonics1,NHphotonics2}, acoustics \cite{NHacoustics}, and cold atomic setups \cite{NHcoldatom}. Another notable feature of the NH systems with open boundary conditions (OBC) is that their wavefunctions are localized at either of the boundaries rather than extending throughout the bulk, a phenomenon also known as the non-Hermitian skin effect (NHSE) \cite{NHSE1,NHSE2,NHSE3,NHSE4,NHSE5,NHSE6,NHSE7,NHSE8,NHSE9}. This results in a significant discrepancy between periodic boundary conditions (PBC) and OBC, leading to the breakdown of the conventional bulk-boundary correspondence (BBC). To address this issue and accurately characterize non-trivial states, various strategies have been proposed, including biorthogonal eigenstates \cite{biorthogonalNHSE}, singular value decomposition \cite{SVDNHSE}, gauge transformations \cite{gaugeNHSE} etc. Further, a key advancement in this area is the emergence of non-Bloch band theory, which extends the traditional Brillouin zone (BZ) to a generalized Brillouin zone (GBZ) in the complex plane \cite{NHSE2,GBZ1,GBZ2,GBZ3,GBZ_SSH_Floquet1,GBZ_SSH_Floquet2,gong_NH_ssh}. This extension allows for the definition of universal topological invariants based on the non-Bloch Hamiltonian, enabling a precise identification of the non-trivial edge states. 
\begin{figure}[t]
    \begin{subfigure}[b]{\columnwidth}
         \includegraphics[width=\columnwidth]{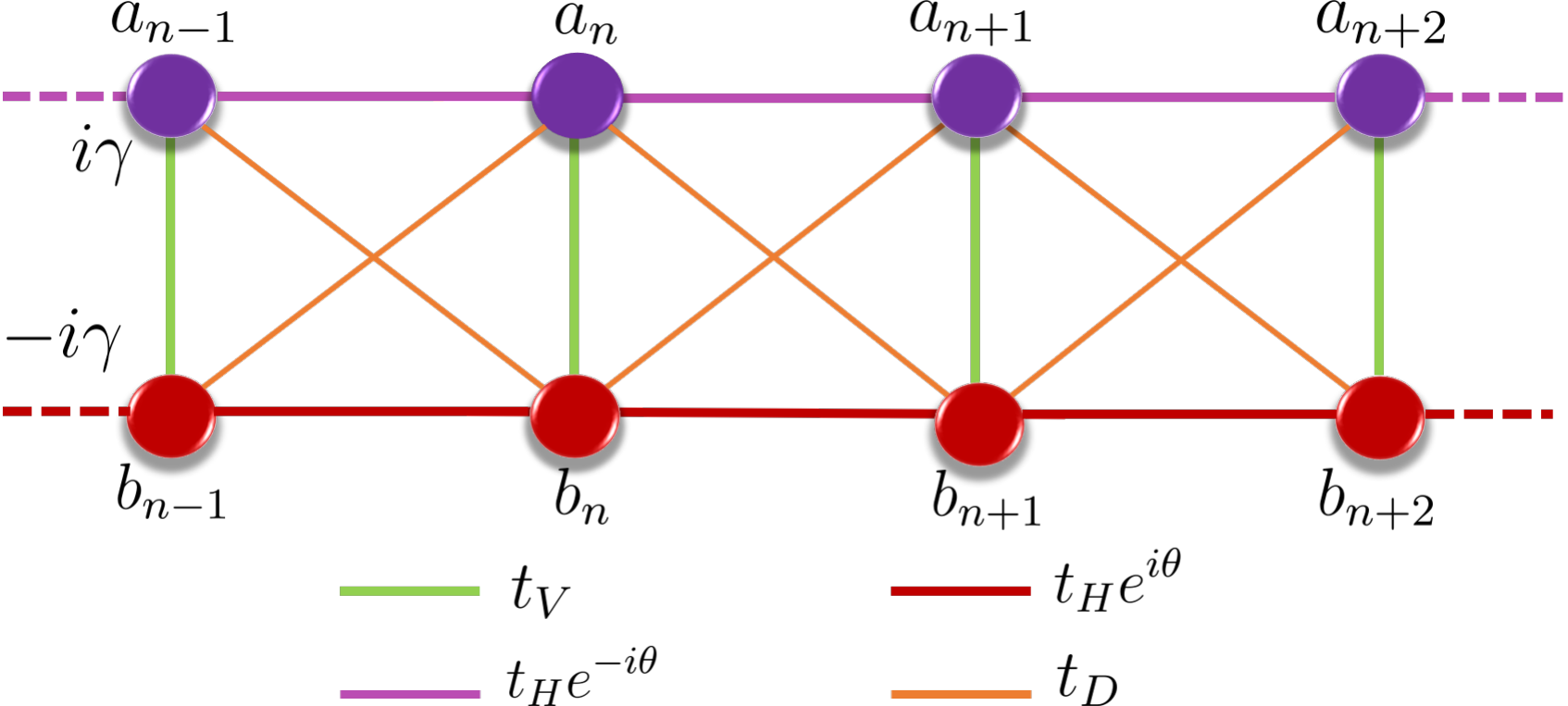}
         \label{1}
     \end{subfigure}
\caption{{The figure depicts a schematic representation of the quasi-1D Creutz ladder, where $a_n$ and $b_n$ denote the two distinct sublattices. The different hopping amplitudes, $t_H$, $t_V$, $t_D$ denote the horizontal, vertical and diagonal hoppings respectively.}} 
\label{Figure_1}
\end{figure}
\par With these realizations, restoring BBC using GBZ has been successfully demonstrated both theoretically \cite{GBZtheory1,GBZtheory2,dipendu3,GBZ_SSH_Floquet3} and experimentally \cite{GBZexpt1,GBZexpt2}. However, most of these studies have been limited to static scenarios. Consequently, it would be highly interesting to explore how non-Hermitian effects interplay with the periodic driving. With this goal in mind, we aim to investigate the Floquet aspects of the NHSE using the Creutz ladder as a prototype. While considerable research has been done on the Creutz ladder in both the static and periodic contexts \cite{NHCreutz1,NHCreutz2}, these studies often did not address the NHSE and hence the GBZ perspective was missing. Therefore, We seek to explore scenarios where the GBZ perspective can be effectively explored. 
\par In light of this, several fundamental questions emerge for us to look into various aspects, such as how does NHSE respond to periodic driving? Is there a directional dependence of NHSE with respect to periodic driving? Can periodic driving eliminate NHSE for specific parameter settings? How do EPs evolve when subjected to the periodic driving? Will multiple topological edge states persist even in the presence of NHSE? Moreover, our primary aim is to develop a generalized Floquet non-Bloch framework that can adapt to any periodic driving protocol. Specifically, We focus on the delta and step drives to highlight key differences in their impact on NHSE, which underscores the versatility of this formalism. Additionally, a concise discussion on the harmonic drive is also included for the sake of completeness, as its distinct computational approach offers valuable insights while reinforcing the robustness of our method. This comprehensive analysis ensures that our framework can be extended to a wide range of driving protocols.
\par The structure of the paper is organized as follows. Sec. \ref{sec:level2} provides an overview of the static NH Creutz ladder, highlighting its key features. Following this, we introduce the Floquet formalism and outline its key aspects. In Sec. \ref{sec:level3a}, we explore the Floquet topological properties induced by NHSE in presence of a delta drive, along with a detailed discussion of the Floquet GBZ formalism to address the breakdown of conventional BBC. For comparison, Sec. \ref{sec:level3b} analyzes a step drive, and Sec. \ref{sec:level3c} examines a harmonic drive, where we highlight the key distinctions between these three driving protocols. Lastly, Sec. \ref{sec:level4} concludes with a summary of our findings.
\\
\section{\label{sec:level2}Non-Hermitian Creutz ladder and Floquet formalism}
The Creutz ladder consists of two rungs of lattice sites that are coupled by diagonal ($t_D$), vertical ($t_V$) and horizontal ($t_H$) hoppings, as shown in Fig.~\ref{Figure_1}. There are two sublattices $a_n$, $b_n$ within each unit cell. The real space Hamiltonian can be written as,
\begin{equation}
\label{Eq1}
\begin{split}
&H_0 = -\sum_n t_{H}(e^{i\theta}a_{n}^{\dagger}a_{n+1} + e^{-i\theta}a_{n+1}^{\dagger}a_{n} +e^{-i\theta}b_{n}^{\dagger}b_{n+1} \\& \qquad+e^{i\theta}b_{n+1}^{\dagger}b_{n}) +t_{D}(a_{n}^{\dagger}b_{n+1}+b_{n+1}^{\dagger}a_{n}+a_{n+1}^{\dagger}b_n\\&\qquad+b_{n}^{\dagger}a_{n+1})+t_{V}(a_n^{\dagger}b_n+b_n^{\dagger}a_n),
\end{split}
\end{equation}
\vspace*{-0.7cm}
\begin{equation}
    H_0 = H_H + H_D + H_V.
\end{equation}
The complex phase, $\theta$ associated with the horizontal hopping leads to a destructive interference, as a consequence of which localization of particles in a certain region of parameter space is observed. Moreover, the Creutz ladder shows a flat band dispersion for the rungless case, $t_V=0$. In an open boundary condition, this leads to the complete localization of states at the edges. Further, in momentum space, the Hamiltonian reads as,
\begin{equation}
\begin{split}
H_{0}(k)= 2t_{H}\cos(k)\cos(\theta)\sigma_0&+2t_{H}\sin(k)\sin(\theta)\sigma_z\\&+(t_V+2t_{D}\cos(k))\sigma_x.
\end{split}
\end{equation}
Here $\sigma_i=x,y,z$ are the Pauli matrices. If $\phi$ denotes the total flux through each plaquette then, $2\theta=\frac{\phi}{\phi_0}$, where $\phi_0$ denotes the magnetic flux quantum. Furthermore, at $\theta= \pm \pi /2$, the Creutz ladder can make a direct resemblance with the SSH model after the following redefinition of the Pauli matrices,
\begin{equation}
    \sigma_y \rightarrow \sigma_z \rightarrow - \sigma_y.
    \label{spin_rotation}
\end{equation}
\par At this point it is important to talk about the symmetries of the model \cite{licharacterization,hughes,juneman}. The model has an inherent inversion symmetry with respect to a horizontal axis that lies symmetrically between the two legs of the ladder. It is expressed by the relation, $\sigma_x H_{0}(k)\sigma_x=H_{0}(-k)$. Furthermore, it possesses a chiral symmetry, $\sigma_y H_{0}(k)\sigma_y=-H_{0}(k)$, which holds only for the values $\theta=\pm \frac{\pi}{2}$. Moreover, despite of the presence of a magnetic flux, the system has an inherent time reversal symmetry given by, $\sigma_x H^{*}_{0}(k)\sigma_x=H_{0}(-k)$. Lastly, a particle-hole symmetry exists in the system for $\theta=\frac{\pi}{2}$ which is defined by, $\sigma_z H^{*}_{0}(k)\sigma_z=-H_{0}(-k)$. 
\par As compared to a simple SSH chain, the Creutz ladder has more tunable parameters. Thus, richer phenomena may emerge when non-Hermiticity is introduced to the different parameters of the Hamiltonian. Specifically, we shall consider the case with sublattice dependent imaginary potential term, $H_{\gamma}$ being added to the Hamiltonian as,
\begin{equation}
    H_{\gamma} = \sum_{n} i \gamma_A a_{n}^{\dagger} a_n + i \gamma_B b_{n}^{\dagger} b_n.
\end{equation}
In a generic sense, such an imaginary on-site potential ($i\gamma$) does not render non-reciprocity to the system and hence does not contribute to the skin effect. However, by introducing staggered on-site potentials ($\gamma_A = -\gamma_B = \gamma$) followed by a re-assignment of the pseudospin operators, as described in Eq. \ref{spin_rotation}, the parameters $t_V$ and $\gamma$ result in the formation of an asymmetric intra-cell hoppings, similar to those in a non-reciprocal SSH model \cite{NHSE1}. In other words, a non-zero value of $\gamma$ will trigger formation of NHSE in the Creutz ladder. Consequently, topological properties associated with in-gap edge modes shall also be affected by NHSE. Furthermore, the inclusion of the NH term introduces certain symmetry modifications \cite{NH10fold,NH10foldfloquet}. As a result, all the symmetry operations previously discussed must be redefined to account for the presence of \( H_{\gamma} \). In Appendix \ref{appendix1}, we have provided a comprehensive analysis of these redefinitions of the symmetry operations, which play a crucial role in defining a suitable invariant for characterizing the NH version of the Hamiltonian.
\par At this stage, the inclusion of periodic drive takes the scenario one step further by notably enhancing the topological characteristics. A general approach to tackle any periodically driven system is to adopt the Floquet formalism which provides a technique to construct an effective time-independent Hamiltonian that captures the stroboscopic evolution of the system through a time evolution,
\begin{equation}
    \hat{U}(T) = \mathcal{T}  \text{exp}[-i \int_{0}^{T} H(t) dt] =  e^{-iH_{\text{eff}}T}.
    \label{evolutionoperator}
\end{equation}
Here $\mathcal{T}$ denotes the time ordering operator and $H_{\text{eff}}$ is the time-independent Floquet effective Hamiltonian.
\par The core idea of Floquet engineering is to generate non-trivial properties even within a statically trivial regime via incorporation of longer-range interactions. As a result, it is reasonable to expect that in the NH version of the driven model, the physics on the GBZ will register significant deviations compared to the case where the driving is missing. Furthermore, in the context of NHSE, we concentrate on three different driving protocols widely studied in literature, namely, delta drive, step drive, and harmonic drive. Each of these driving mechanisms has distinct effects on NHSE, which we have briefly summarized along with an in-depth analysis provided in the next section. With this in mind, the summarization of our formalism can pave the way for the realization of intriguing NH features through any form of periodic modulation, offering new avenues for exploring non-Hermitian effects in driven systems.

\begin{figure*}[t] 
\begin{minipage}{\linewidth}
        \begin{subfigure}[t]{0.95\columnwidth}
            \includegraphics[width=\linewidth]{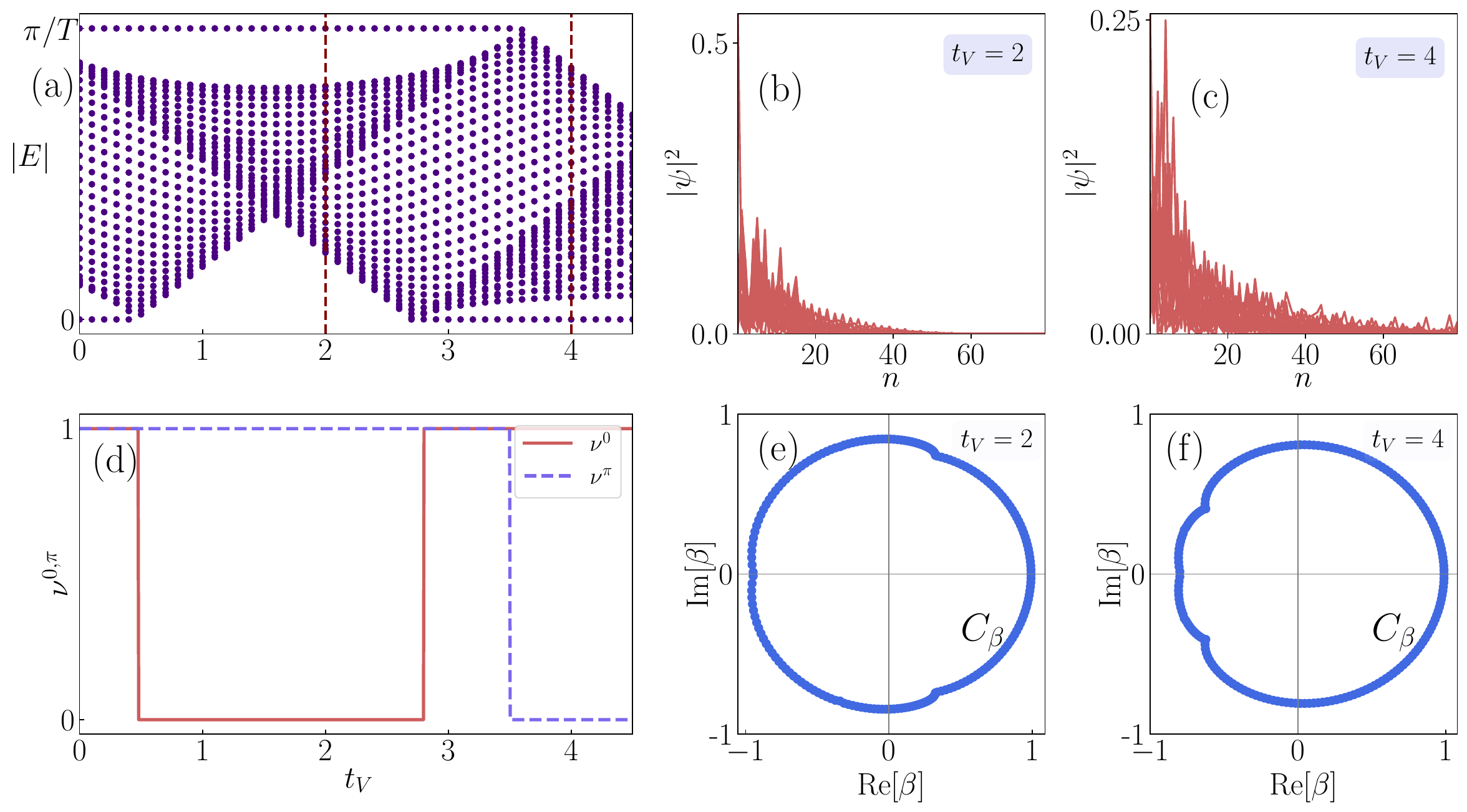}
        \end{subfigure}%
\caption{Panel (a) shows the Floquet quasi-energy spectrum corresponding to the delta drive, plotted as a function of the $t_V$. Panels (b) and (c) describe the probability distributions of the eigenstates, and their corresponding GBZ ($C_\beta$) has been shown in panels (e) and (f). The coordinates of GBZ have been utilized to compute $\nu^{0,\pi}$ as shown in panel (d) which correctly coincides with panel (a). The rest of the parameters are chosen as, $T=1$, $t_H=0.5$, $V_0=\pi/2$, $\gamma=0.4,$ and $t_D=1$.}
\label{Figure_2}
    \end{minipage}
\end{figure*}
\section{\label{sec:level3}Results and Discussions}
\subsection{\label{sec:level3a}Delta drive}
At first, we shall discuss the implications on Floquet topological aspects when the NH Creutz ladder is periodically perturbed by a delta drive which manifests through the vertical hopping ($t_V$) given as,
\begin{equation}
H_V(t)=\Big[ t_V + V_0 \sum_{m=-\infty}^{m=\infty} \delta(t-mT) \Big] \sum_{n} (a_n^{\dagger}b_n+a_nb_n^{\dagger}),
\end{equation}
where $V_0$ is the strength of the drive. The stroboscopic evolution corresponding to a delta drive thus leads to an effective Floquet Hamiltonian that can be obtained from the product of two exponential matrices given as,
\begin{equation}
\label{E25}
\begin{split}
    U(T)= e^{-iV_{0}\sum_{n} (a_n^{\dagger}b_n+a_nb_n^{\dagger})} e^{-iH_0T}  =e^{-iH_{\text{eff}}T},
\end{split}
\end{equation}
In Fig. \ref{Figure_2}a, we present the quasienergy spectrum as a function of $t_V$, with the other parameters set to $t_H=0.5$, $V_0=\pi/2$, $\theta=\pi/2$, $\gamma= 0.4$, $t_D=1$, and $T=1$. For consistency, we use this particular set of parameters throughout this section. Additionally, the energy has been computed in units of $t_D$. Moreover, owing to the chiral symmetry, in addition to the conventional zero energy modes, the system manifests $\pi$ energy modes at the boundary of FBZ. Additionally, in Figs. \ref{Figure_2}b and \ref{Figure_2}c, we show the probability distribution of the eigenstates with OBC for a lattice size $L = 2N = 80$. Eventually, all the eigenstates get localized at one particular end of the system, which we refer to as uni-directional NHSE \cite{GBZtheory1}. This localization prevents us from characterizing the edge states by simply analyzing the topology of the bulk, resulting in a breakdown of the conventional BBC. Nevertheless, we can restore this correspondence based on the framework of non-Bloch band theory via formulating a generalized version of BZ (GBZ). This approach allows us to map the open boundary spectra from the continuum bands and predict the emergence of zero and $\pi$ energy modes using non-Bloch invariants.
\par Before delving into the precise evaluation of topological invariants using non-Bloch theory, it is important to first highlight the critical challenges introduced by the application of a periodic drive. Notably, sublattice symmetry (SLS), analogous to the chiral symmetry (CS) in the Hermitian limit, plays a pivotal role in identifying non-trivial phases through the computation of non-Bloch band invariants. However, periodic driving often disrupts these symmetries, as the Hamiltonian at different times generally fails to commute. One can reconcile this fact by expressing the coefficients of $H_{\text{eff}}(k)$ in the momentum space,
\begin{equation}
    H_{\text{eff}}(k) = d_x \sigma_x + d_y \sigma_y + d_z \sigma_z,
    \label{all_d_componets}
\end{equation}
(See Appendix \ref{appendix} for more details on the explicit expression of each $d$-vectors). From this, it becomes evident that $H_{\text{eff}}(k)$ typically includes three components, whereas the chiral symmetric $H(k)$ contains only two. This demonstrates that periodic driving can indeed break CS as well as SLS in the NH limit. While the method for restoring broken symmetries in Floquet-driven systems is well-established in the literature, we intend to offer a detailed pedagogical explanation of the method, as the explicit expressions are essential for computing the GBZ. 
\par To begin with, it is interesting to note that different choices of the time period can result in the effective Hamiltonian ($H_{\text{eff}}$) possessing different symmetries or none at all. As a remedy, we rely on the mechanism of pair of `\textit{symmetric time frames}' \cite{asbothwinding1,asbothwinding2} that are defined by the specific choice of time frames, resulting in the Floquet evolution operator assuming a form,
\begin{equation}
    \hat{U} = \hat{F} \hat{G},
\end{equation}
where, $\hat{F}$ and $\hat{G}$ in the Hermitian limit are related by the chiral symmetry operator as,
\begin{equation}
    \hat{\Gamma} \hat{F^{\dagger}} \hat{\Gamma} = \hat{G}^{-1}.
    \label{chiral_partner}
\end{equation}
It is also easy to verify that if a symmetric time frame exists corresponding to a Floquet evolution operator $\hat{U}_{1}$ = $\hat{F} \hat{G}$, then there must also exist another symmetric time frame corresponding to the Floquet operator $\hat{U}_{2}$ = $\hat{G} \hat{F}$. Now, let us consider that the Floquet operator in one of the time frame from $t = T/2$ to $3T/2$ read as,
\begin{equation}
    U_{1,k}  =e^{-iH_0(k)T/2} e^{-iV_{0}\sigma_x}  e^{-iH_0(k)T/2} =e^{-iH_{\text{1,eff}}T},
\end{equation}
Similarly, using the chiral symmetry operator, the Floquet time evolution in the second symmetric time frame assumes the form,
\begin{equation}
\begin{split}
    U_{2,k} =e^{-iV_{0}\sigma_x /2} e^{-iH_0(k)T} e^{-iV_{0}\sigma_x/2} =e^{-iH_{2,\text{eff}}T}.
\end{split}
\end{equation}
Both $H_{j,\text{eff}}$ (where $j=1,2$ denotes an index for the frames used) are two-component vectors, and can be written as,
\begin{widetext}
\begin{subequations}
\begin{align}
d_{1x} &= \sin(V_0) \left[ \cos(E_0 T) \left(\frac{d_x}{E_0}\right)^2 + 1 -  \left(\frac{d_x}{E_0}\right)^2 \right] +  \left(\frac{d_x}{E_0}\right) \left[ \sin(E_0 T) \cos(V_0) \right], \\
d_{1z} &= \sin(V_0) \left[ \frac{d_x d_z}{E_0^2} \left( \cos(E_0 T) - 1 \right) \right] + \left(\frac{d_z}{E_0}\right) \left[ \sin(E_0 T) \cos(V_0) \right],
\end{align}
\label{d_vectors_delta_drive_1}
\end{subequations}

\begin{equation}
d_{2x} = \cos(V_0) \sin(E_0 T) \left( \frac{d_x}{E_0} \right) + \sin(V_0) \cos(E_0 T), \qquad 
d_{2z} = \sin(E_0 T) \left( \frac{d_z}{E_0} \right),
\label{d_vectors_delta_drive_2}
\end{equation}
\end{widetext}
(see Appendix \ref{appendix} for the detailed derivation of the $d$-vectors) where, $d_x = t_V + 2t_D \cos(k) $, $d_z = 2t_H \sin(k) - i \gamma$, and $E_0 = \sqrt{d_{x}^2 + d_{z}^2}$ is the eigenvalue of the static Hamiltonian. Although, neither of the frames gives complete information about the number of edge modes. Rather, based on the periodic table of Floquet topological insulators \cite{roy10fold} each of the nontrivial phases of the system can be characterized by a pair of winding numbers $\nu^{0}$ and $\nu^{\pi}$, given by,
\begin{equation}
    \nu^0 = \frac{\nu_1 + \nu_2}{2} \quad ; \quad \nu^{\pi} = \frac{\nu_1 - \nu_2}{2}.
\label{symmetric_frame}
\end{equation}
Here, $\nu_1$ and $\nu_2$ are the winding numbers for the two effective Hamiltonians corresponding to the two symmetric time frames. Furthermore, in the context of 1D systems with restored SLS, the Hamiltonian is effectively classified in the first homotopy group \cite{10fold,schnyder}, providing us with the winding number as the topological invariant which is defined as,
\begin{equation}
\nu_j = \frac{1}{2\pi i} \int_{C_k} L_j^{-1}(k) \frac{d}{dk} L_j(k) dk,
\label{winding_number}
\end{equation}
where, $L_j(k) = d_{j,x}(k) + i d_{j,z}(k)$ and $C_k$ stands for the conventional BZ ($k \in [0:2\pi$]). To compute these invariants \((j = 1, 2)\), it is necessary to perform integration over an appropriate GBZ, which plays a vital role in recovering the disrupted BBC. In the next section, we will introduce a Floquet non-Bloch framework that offers a comprehensive method for determining the GBZ in the context of periodically driven systems.
\subsubsection{\label{GBZ}\textbf{Generalized Brillouin zone and $\mathbb{Z} \times \mathbb{Z}$ invariants}}
The precise description of the topological characterization necessitates restoring BBC even in the presence of NHSE, which can be challenging due to the significant differences between PBC and OBC spectra, an artifact of NHSE. To address this, we utilize the non-Bloch theory within the GBZ framework. This approach involves expressing the Floquet Bloch Hamiltonian in terms of a quantity $\beta=e^{ik}$ ($k\in [-\pi:\pi])$, leading to the non-Bloch Floquet Hamiltonian $H(\beta)$. one can determine $\beta$ using the characteristic polynomial $\det[H(\beta)-E]$ = 0, which yields an algebraic equation for $\beta$ of even degree. Further, if the solutions are numbered as $\beta_i$ $(i=1,2,...2N)$ and satisfy $|\beta_1|\le|\beta_2|\le|\beta_3|\le ... \le|\beta_{2N}|$, then GBZ ($C_{\beta}$) can be determined by tracing the trajectory of $\beta_N$ and $\beta_{N+1}$, ensuring the condition, $|\beta_N| = |\beta_{N+1}|$.
\par This method has been previously demonstrated for several static systems \cite{GBZtheory1,GBZtheory2}. However, applying the GBZ to Floquet systems introduces technical challenges owing to mathematical complexities. For instance, in static systems, the effective Hamiltonian generally features only short-range interactions, leading to the solution of the characteristic polynomial, $\det[H(\beta) - E] = 0$ of finite order in $\beta$. In contrast, periodic driving even in our simple quasi-1D Floquet system introduces longer-range interactions, resulting in a significantly complicated characteristic polynomial. 
\begin{figure}[t]
\hspace{-0.9cm}
    \begin{subfigure}[b]{\columnwidth}
         \includegraphics[width=\columnwidth]{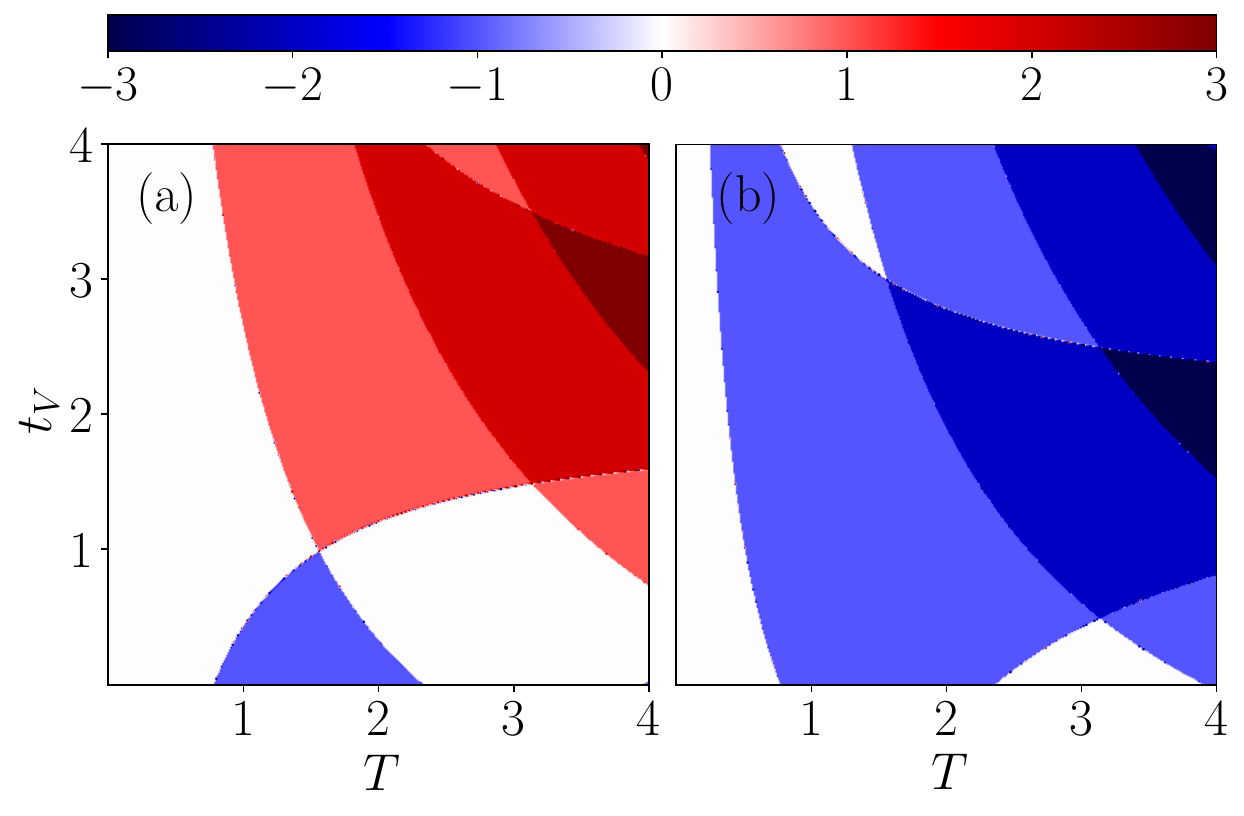}
         \label{1}
     \end{subfigure}
\caption{{The figure depicts the topological phase diagram in $t_V-T$ plane computed using the winding numbers corresponding to the zero ($\nu^{0}$) modes in panel (a) and the $\pi$ ($\nu^{\pi}$) modes in panel (b). The rest of the parameters are chosen as, $t_H=0.5$, $V_0 = \pi/2$ and $\gamma=0.4$.}} 
\label{Figure_3}
\end{figure}
Moreover, the involvement of such longer-range interactions in a Floquet scenario can be observed from the momentum space functions, $\sin(F(\cos{k}))$ in Eqs. \ref{d_vectors_delta_drive_1} and \ref{d_vectors_delta_drive_2}, which directly reflect these long-range interactions across the system. As a result, the characteristic equation might involve terms that are infinite-order polynomials in $\beta$. To address this, one can assume that the hopping strengths (such as $t_H$ or $t_D$) associated with $\beta$ is less than 1, allowing the truncation of the polynomial expansion in $\beta$ upto a finite order, say $\beta^{n}$ (and $\beta^{-n}$), using an $n^{\text{th}}$ order Taylor expansion. Moreover, it can be observed that as long as \( t_H \leq 0.5 \), truncating the polynomial to the second order results in a reliable approximation. In Appendix \ref{appendix3}, we offer a numerical validation that supports the correct truncation order for the characteristic polynomial. This simplification reduces the characteristic equation to yield quartic solutions for $\beta$.
\begin{figure}[t]
\hspace{-0.8cm}
    \begin{subfigure}[b]{\columnwidth}
         \includegraphics[width=0.95\columnwidth]{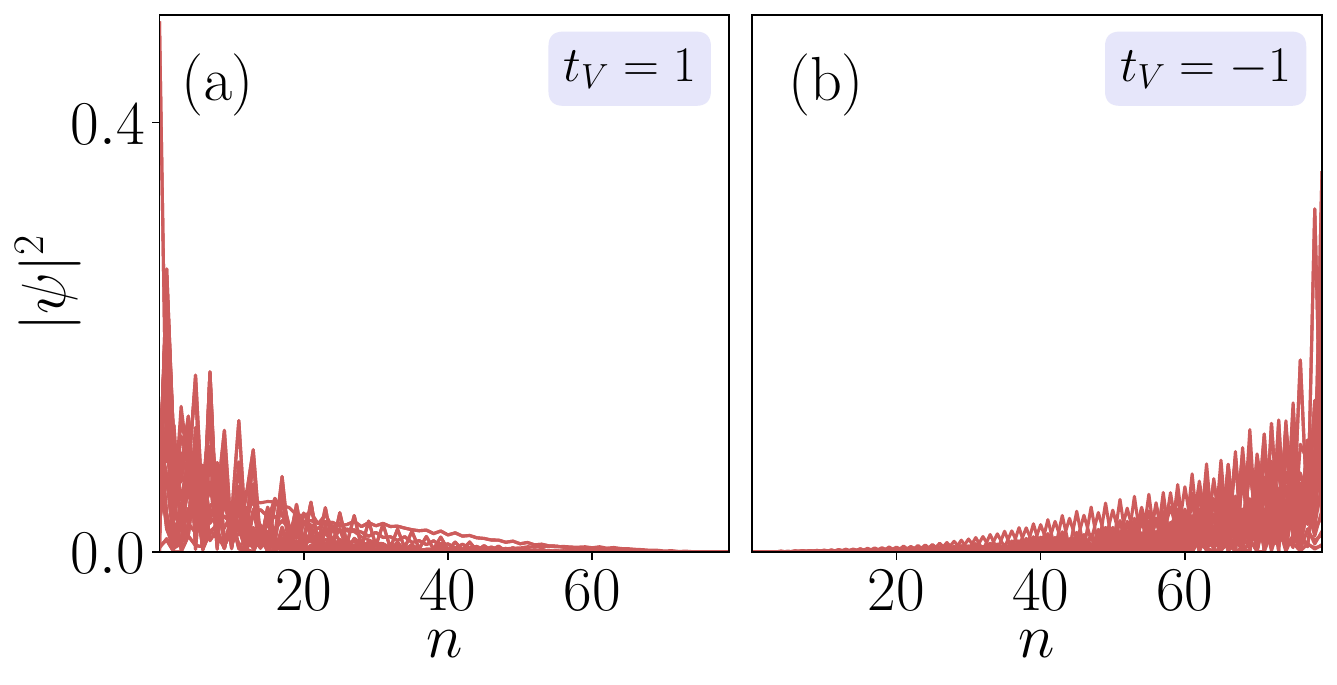}
         \label{1}
     \end{subfigure}
     \begin{subfigure}[b]{\columnwidth}
         \includegraphics[width=1\columnwidth]{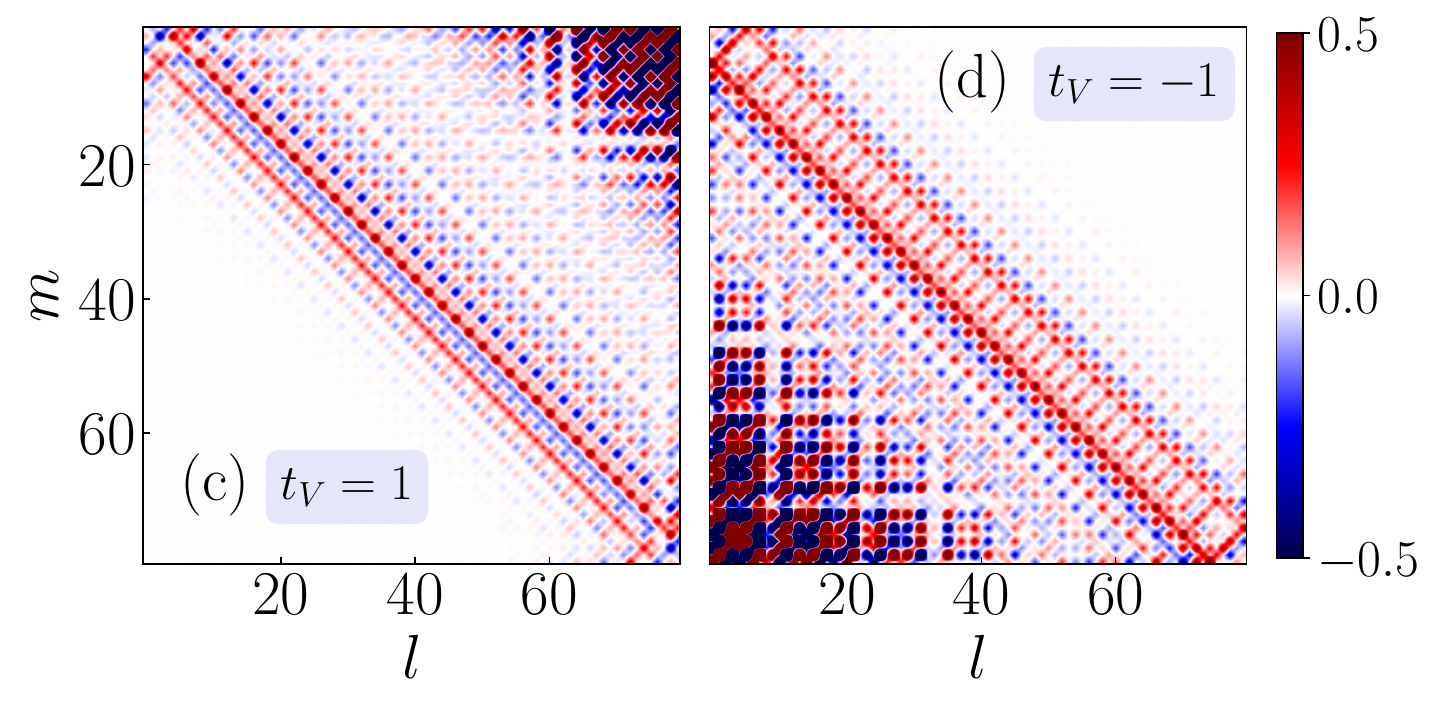}
         \label{1}
     \end{subfigure}
\caption{{Panels (a) and (b) denote the distribution of the skin modes at the edges corresponding to positive ($t_V=1$) and negative ($t_V=-1$) values of $t_V$ respectively. The associated expansion coefficients corresponding to the Hamiltonian for both $t_V=1$ and $t_V=-1$ in the coordinate basis have been shown in panels (c) and (d) respectively. Here $l$ and $m$ denote the base indices. The other parameters have been chosen as $t_H=0.5$, $V_0=\pi/2$, $T=1$, $\gamma=0.5$, and $t_D=1$.}} 
\label{Figure_4}
\end{figure}
\par We begin by replacing $e^{ik}$ by $\beta$, enabling us to rewrite the characteristic polynomial in terms of the $d$-vectors given as,
\begin{equation}
    E^2(\beta) = d_{j,x}^2(\beta) + d_{j,z}^2(\beta); \quad (j=1,2).
    \label{characteristic_eqn_d_vectors}
\end{equation}
The selection of $\beta$ must ensure that the energy levels are dense enough to asymptotically form continuum bands. Additionally, to obtain an approximate analytical solution, a Taylor expansion by considering small $t_H$ (with the condition \( t_H = 2t_D \), such that for \( t_H = 0.5 \), \( t_D \) becomes 1, allowing for the verification of the results shown in Fig. \ref{Figure_2}) is carried out up to a specific order. As a result, Eq. \ref{characteristic_eqn_d_vectors} can be simplified to,
\begin{equation}
    E^2(\beta) = \sum_{j=-2}^{j=2} \beta_j X^j,
    \label{Taylor}
\end{equation}
where $X^{j}$ are the coefficients of $\beta_j$. Note that, the degree of the polynomial depends on the order of the Taylor expansion. Now assume there are two solutions $\beta$ and $\beta^{\prime} (= \beta e^{i\theta}$) that share the same absolute value. Therefore, subtracting the characteristic equations, $E^{2} = F(\beta)$ and $E^{2} = F(\beta^{\prime})$ would result in,
\begin{equation}
    0 = \sum_{j=-2}^{j=2} \beta_j X^j ( 1 - e^{ij\theta} ),
    \label{Taylor2}
\end{equation}
which allows us to determine the solutions of $\beta$ over a specific range of $\theta$, say $\theta \in [0:2\pi]$. Finally, we arrange all the $\beta$ solutions in ascending order based on their absolute values, and by selecting those that satisfy the condition $|\beta_{2N}| =|\beta_{3N}|$(with 
$N$ being the number of intervals within the $\theta$ range), we ultimately arrive at the GBZ which we call as $C_{\beta}$.
\par In Figs. \ref{Figure_2}e and \ref{Figure_2}f, we present the GBZ for two specific points, as indicated in Fig. \ref{Figure_2}a, corresponding to $t_V =2$ and $t_V =4$, respectively. It is evident that the GBZ always deviates from the conventional BZ, which leads to the emergence of NHSE. Moreover, the GBZ exhibits a similar structure in two symmetric frames, validating the accuracy of our method. Additionally, $C_{\beta}$ may feature cusps where multiple solutions of $\beta$ share the same absolute value. The recovery of BBC can be verified by calculating the generalized non-Bloch invariants as outlined by Eq. \ref{symmetric_frame} and \ref{winding_number}, where $C_k \in [0:2\pi]$ has to be replaced with $C_{\beta}$ in order to accurately determine the topological phase transitions. Additionally, for computational convenience, we set the driving strength as $V_0 = \pi/2$ (as said earlier), allowing the elimination of the second term in the $x$ and $z$-components for the first frame. This simplifies the method, enabling a more straightforward access to the Taylor expansion as shown in Eq. \ref{Taylor}.
\begin{figure}[t]
    \begin{subfigure}[b]{\columnwidth}
         \includegraphics[width=0.9\columnwidth]{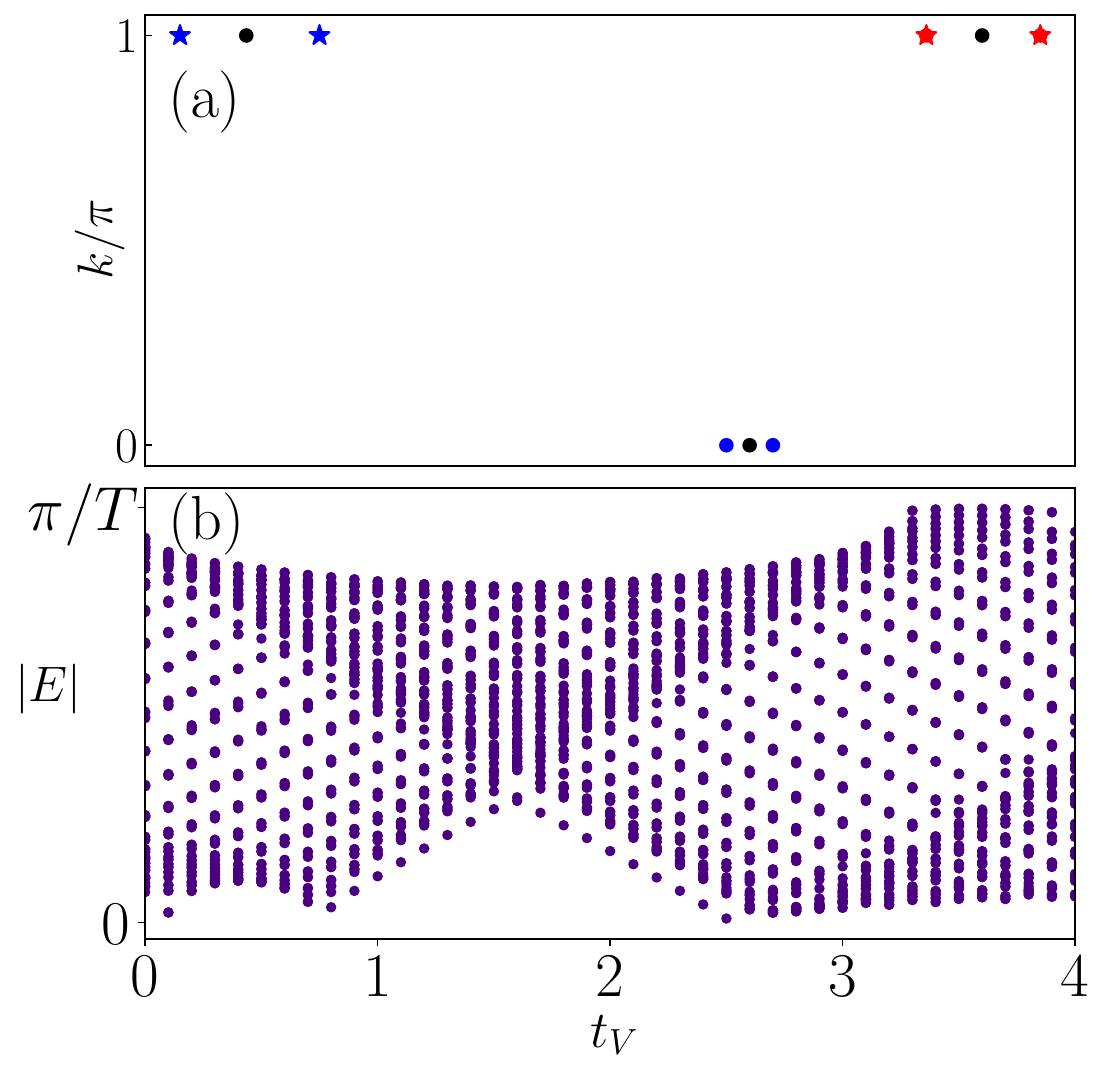}
         \label{Figure_5}
     \end{subfigure}
\caption{{Panel (a) depicts the location of the EPs that demarcates the quasienergy gap closure at $E=0$ with blue and $E=\pi$ with red labels. It is seen that each of the DPs, marked by a black circle, splits into two EPs due to the presence of non-Hermiticity, with each EP inheriting half of the topological winding number from the original DP.  Additionally, gap closures at the edges of the FBZ are marked with a star, while those at the centers are indicated by a circle. The emergence of EPs precisely corresponds to the PBC spectrum shown in panel (b). The rest of the parameters are chosen as, $t_H=0.5$, $V_0=\pi/2$, $T=1$, $\gamma=0.4$, and $t_D=1$.}} 
\label{Figure_5}
\end{figure}
\par In Fig. \ref{Figure_2}d, we present the non-Bloch invariants, $\nu^{0,\pi}$ as a function of $t_V$, which accurately capture the emergence of zero and $\pi$ energy modes, as depicted in Fig. \ref{Figure_2}a. Additionally, we have plotted the phase diagram in the $t_V - T$ plane for both the zero and $\pi$ winding numbers, as illustrated in Fig. \ref{Figure_3}. Our findings reveal emergence of higher winding numbers in the low-frequency regime. The presence of such non-trivial phases originates from the distinguished role of periodic driving in simulating an effective long-rane interactions. The way one can reconcile this is to do a Baker-Campbell-Hausdorff expansion where the nested commutators that induce coupling between distant sites grow as a function of $T$ which effectively yields more off-diagonal terms in $\hat{U}(T)$, therefore, resulting in the generation of multiple edge states.
\par Furthermore, it is fascinating to note that, depending on the sign of $t_V$, all the elements in $H_{\text{eff}}$ corresponding to either the upper or the lower diagonal sites can become fully occupied, leaving the rest completely empty. This phenomenon results from the interplay between periodic driving and non-Hermiticity, where the imaginary on-site potential effectively induces non-reciprocity among the hopping amplitudes, and this effect gets further intensified by the application of the periodic drive. Fig. \ref{Figure_4} clearly demonstrates that when $t_V$ is positive (negative), all the elements in $H_{\text{eff}}$ corresponding to the upper (lower) diagonal are fully occupied, leaving the rest completely empty. As a result, all the eigenstates are localized at the left (right) end of the system. This observation suggests that $t_V = 0$ can be considered a critical point for the NHSE. Furthermore, it is crucial to recognize that there is always a simultaneous competition between NHSE-induced localization and topological localization. While NHSE causes uni-directional localization of all eigenstates, topological localization results in bi-directional localization of edge states only. Therefore, at $t_V = 0$, topological localization dominates, and if the system with $t_V =0$ falls in a topologically non-trivial phase, it can exhibit a pair of edge modes localized at both the ends, along with the bulk skin modes induced by NHSE. Another remarkable feature of GBZ is that when all the eigenstates are localized at left (right) edge of the system the corresponding $C_{\beta}$ has a radius lesser (greater)  than 1.
\begin{figure}[t]
    \begin{subfigure}[b]{\columnwidth}
         \includegraphics[width=0.9\columnwidth]{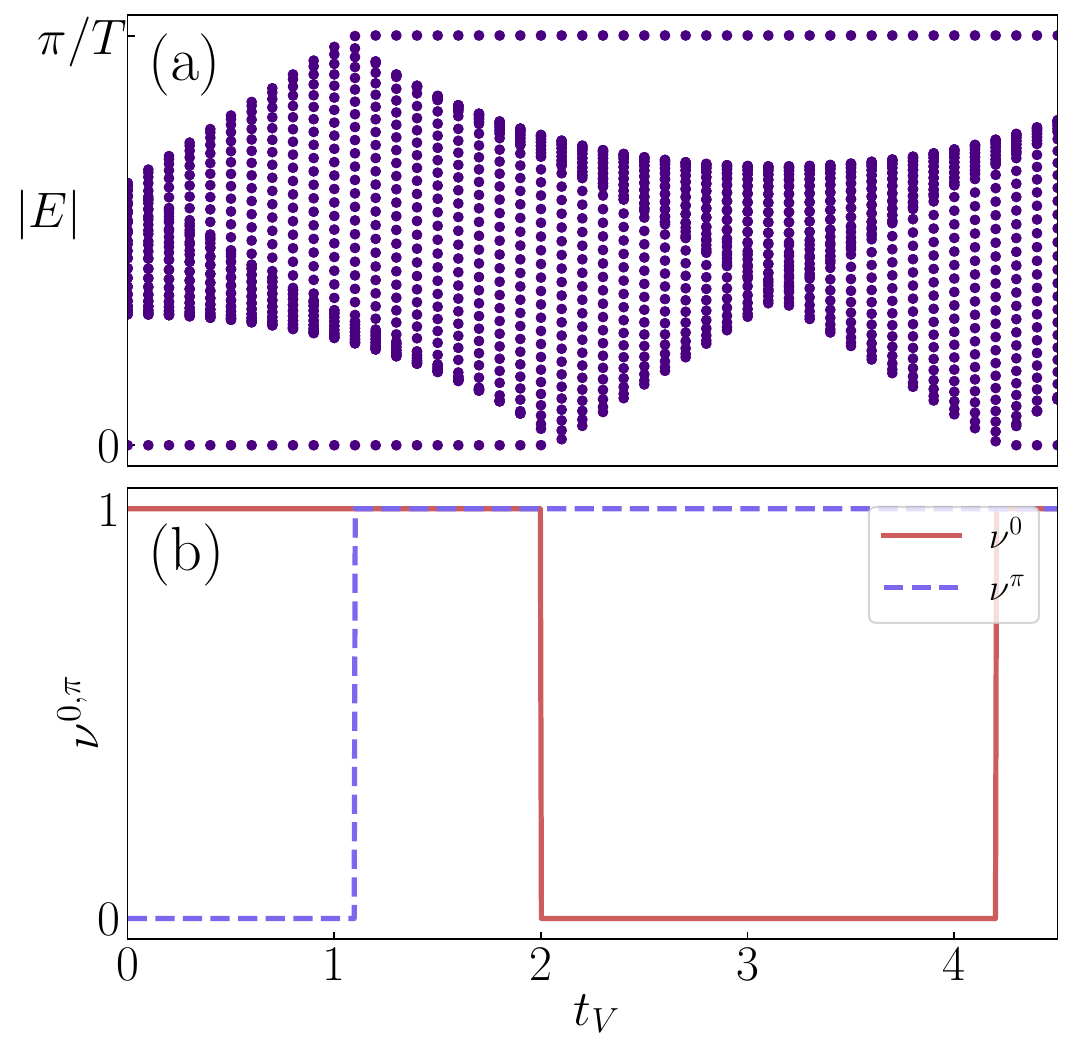}
         \label{1}
     \end{subfigure}
\caption{{Panel (a) shows the Floquet quasi-energy spectrum corresponding to the step drive, plotted as a function of the $t_V$. Panel (b) depicts the variation of $\nu^{0,\pi}$ as a function of $t_V$ with the other parameters being chosen as, $t_H=0.5$, $T=2$, $\gamma=0.4$, and $t_D=1$.}} 
\label{Figure_6}
\end{figure}
\subsubsection{\textbf{Exceptional points}}
In the context of NHSE, where eigenstates are localized at the boundaries, the exceptional points (EPs) serve as critical markers as they often signal transitions between different phases. EPs are significant as they may disrupt BBC, leading to changes or ambiguities in the topological invariants, such as the winding number. Generally, EPs arise when all the eigenvalues and eigenvectors coalesce. Unlike Hermitian systems, where the eigenstates remain orthogonal at band-touching degenerate points (DPs), EPs cause the eigenstates to lose orthogonality, and the Hamiltonian becomes non-diagonalizable. Consequently, it is reasonable to expect that at EPs, both the real and the imaginary parts of the eigenvalues approach zero. Furthermore, Floquet topological phase transitions, marked by the zero and the $\pi$ energy modes, are associated with the closure of quasi-energy bands at exceptional points (EPs) where $E(k) = 0, \pm \pi$. With the expression of $E(k)$ being,
\begin{equation}
\begin{split}
    E_k = \frac{1}{T} \arccos & \Big[ \cos V_0 \cos(E_{k,0}T) \\ & - \frac{t_V + 2t_D \cos k}{E_{k,0}} \sin V_0 \sin (E_{k,0}T)\Big],
\end{split}
\end{equation}
(See Appendix \ref{appendix} for the derivation of $E(k)$).
Moreover, for a driving strength of $V_0 = \pi/2$, the locations of these EPs can be identified by solving,
\begin{equation}
   \frac{t_V + 2t_D \cos k}{E_{k,0}} \sin V_0 \sin (E_{k,0}T) = \pm 1.
\end{equation}
In Fig. \ref{Figure_5}a, we present the location of EPs for different quasi-energies as a function of $t_V$ based on the parameter choices outlined in Sec. \ref{sec:level3a}. These results are consistent with the periodic boundary spectrum depicted in Fig. \ref{Figure_5}b. Furthermore, it is noteworthy that the EPs can be thought of as a NH analog of the DPs. This is due to the fact that non-Hermiticity may cause a single DP to split into two EPs. Each pair of EPs, indicated by red or blue circles (when the bulk gap closes at the center of FBZ)/stars (when the bulk gap closes at the edges of FBZ), originates from a DP (in black) as illustrated in Fig. \ref{Figure_5}a.  
In principle, each EP then inherits half of the topological invariant of the DP associated with the parent Floquet Hermitian system. Consequently, the region between the two EPs delineates an NH-induced phase with fractional winding numbers yet does not necessarily forecast the presence of localized edge states.
Moreover, as the NH term strengthens, the gap between the EPs widens, leading to a more serious disruption of BBC. This underscores the need for constructing the GBZ, which may account for the complex topology introduced by the EPs, and can accurately reproduces the features of the OBC, even in the presence of a strong $\gamma$. Additionally, a recent study \cite{gong_NH_ssh} has demonstrated the discovery of a novel class of EPs in two-dimensional models, which can emerge independently of their Hermitian counterparts.
\subsection{\label{sec:level3b}Step-Drive}
In this section, we explore the effects of Floquet driving on the NHSE when the system is subjected to a step drive. Similar to the delta drive applying the Floquet formalism requires constructing the time evolution operator, which once again involves the product of two exponential matrices. Yet, the step drive introduces unique NH features that are absent in the delta drive, thus highlighting the need for its separate investigation.
\par Let us consider the step drive in terms of periodic quenches given as,
\begin{equation}
   H(t) = 
\begin{cases} 
H_1, & \text{for } t \in [lT, lT + T/2) \\
H_2, & \text{for } t \in [lT + T/2, lT + T)
\end{cases} 
\end{equation}
where $j \in \mathbb{Z}$ and $T$ is the time driving time period. Within each period, the Hamiltonians, $H_1$ and $H_2$ are given as,
\begin{equation}
    H_1 = H_H + H_{\gamma} \quad ; \quad H_2 = H_D + H_V
\end{equation}
The Floquet operator for this periodically quenched model, which governs the dynamics over a complete period $T$, that is, from $t= jT + 0^{-}$ to $(j+1)T + 0^{-}$ is then given as,
\begin{equation}
    U(k) = e^{-i d_x (k) \sigma_x T/2} e^{ - i d_z (k) \sigma_z T/2}.
    \label{evolution_step_drive}
\end{equation}
In Fig. \ref{Figure_6}a, we have plotted the quasienergy spectra corresponding to $H_{\text{eff}}$ for the specific choice of parameters namely, $t_H=0.5$, $t_D=1$, $\gamma=0.4$ ,and $T=2$. It is worth noting that the Floquet effective Hamiltonian can exhibit NHSE even when the quenched Hamiltonians at each step do not exhibit NHSE.
\begin{figure*}[t] 
\begin{minipage}{\linewidth}
        \begin{subfigure}[b]{0.95\columnwidth}
            \includegraphics[width=\linewidth]{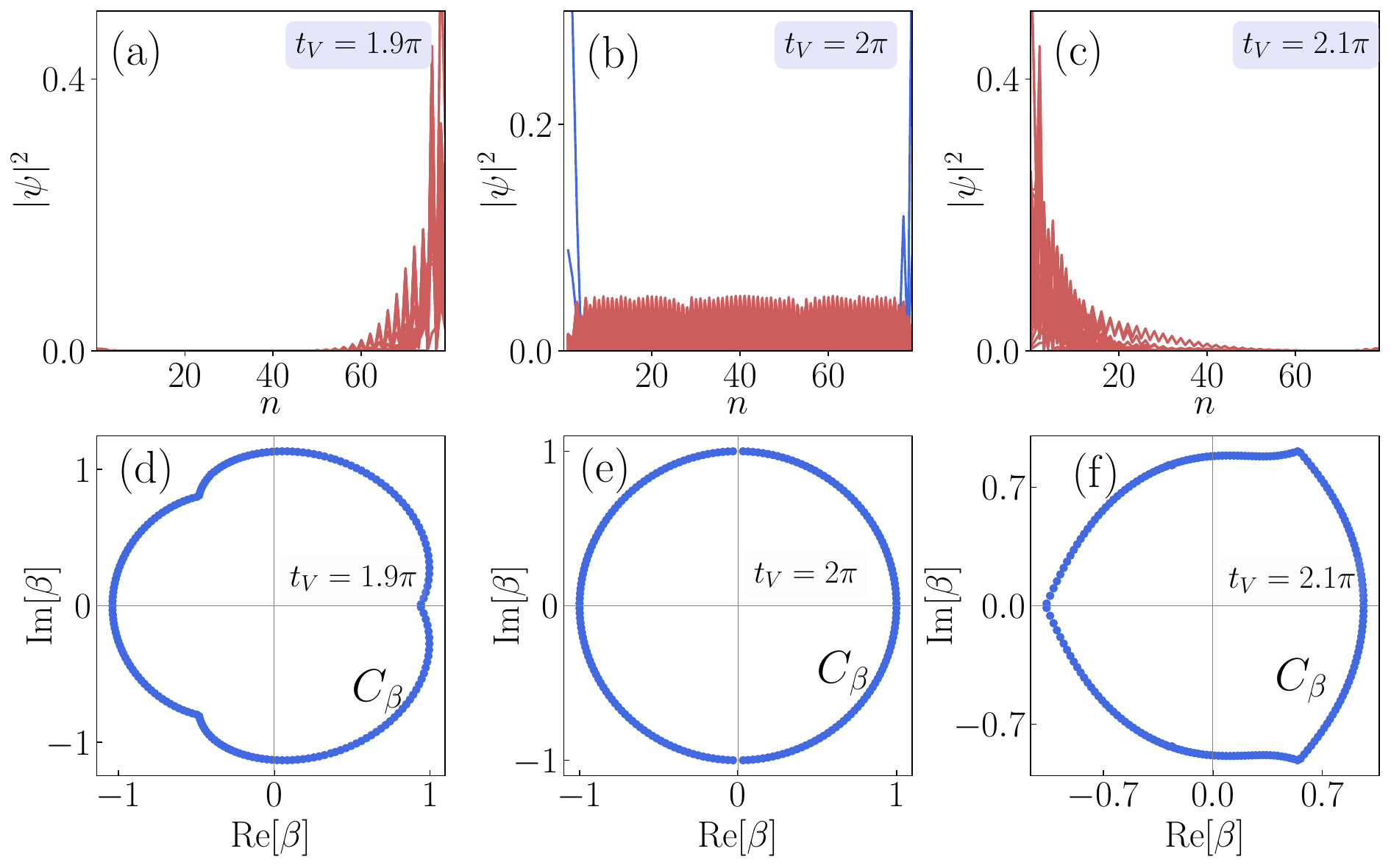}
        \end{subfigure}%
\caption{Panel (a) and (c) represent the localization of the skin modes at the right and left edges of the system, and their corresponding GBZ with their radius being greater and lesser than 1 are shown in panel (d) and panel (f) respectively. While in panel (b) the probability distribution reveals that the system is free from NHSE at $t_V = 2\pi$. Consequently, the corresponding GBZ with unit radius has been shown in panel (e). The rest of the parameters are chosen as, $t_H=0.5$, $T=2,$ $\gamma=0.4$, and $t_D=1$.}
\label{Figure_7}
    \end{minipage}
\end{figure*}
\subsubsection{\textbf{Generalized Brillouin zone and $\mathbb{Z} \times \mathbb{Z}$ invariants}}
Similar to the delta-driven case, the GBZ for the step drive can be derived by conducting a second-order Taylor expansion of the characteristic polynomial about $t_H=0$, followed by a substitution of $e^{ik}$ with $\beta$. Applying similar mathematical techniques as earlier, the $d$-vectors for the two symmetric frames can be derived as,
\begin{subequations}
\begin{align}
d_{1x} &= \sin(d_x) \cos(d_z), \quad d_{1z} = \sin(d_z), \\
d_{2x} &= \sin(d_x), \qquad \qquad d_{2z} = \cos(d_x) \sin(d_z).
\end{align}
\label{d_vectors_step_drive}
\end{subequations}
See Appendix \ref{appendix} for detailed derivation of the $d$-vectors where for the simplicity of calculations, we have fixed $T=2$. By substituting these $d$-vectors into the characteristic polynomial and solving the resulting equation, the trajectory of the GBZ can be determined with the boundary condition $|\beta_{2N}| = |\beta_{3N}|$. Just like in the delta drive case, the GBZ does not uphold a uniform radius and instead displays cusps.
\par However, the step-driven model reveals intriguing features that make it distinct from the delta-driven case. For instance, at specific periods, $T = 2n$ ($n \in \mathbb{Z}$), and parameter values, such as $t_V = n\pi$ the system can be free of NHSE. This indicates that $t_V = n\pi$ acts as a critical boundary that marks a transition in the behavior of NHSE. For example, at $t_V = 1.9 \pi$, all the eigenstates are localized at the right end of the system (Fig.~\ref{Figure_7}a), and the corresponding GBZ has a radius greater than 1 (Fig.~\ref{Figure_7}d). As $t_V$ increases to $t_V = 2\pi$ (with $n=2$), the system becomes free of NHSE. At this point, the eigenstates are extended across the bulk with two localized edge modes (Fig.~\ref{Figure_7}b). This behavior is similar to that seen in the Hermitian version of the model, which is why we refer to these points as the Floquet Hermitian points. Additionally, the GBZ at $t_V = 2\pi$ takes the form of a unit circle, indicating that at these points the GBZ has transformed into a conventional BZ (Fig.~\ref{Figure_7}e). When $t_V$ is increased further, for example at $t_V = 2.1 \pi$, the system exhibits NHSE again, however, at this value, all eigenstates get localized at the left end (Fig.~\ref{Figure_7}c), and the corresponding GBZ has radius lesser than 1 (Fig.~\ref{Figure_7}f). These findings reveal an intriguing prospects in the theory of quantum information, where NHSE-induced localization through periodic driving enables instantaneous edge-to-edge communication even when the boundaries are arbitrarily far from each other. This precise control of NHSE via periodic modulation may hold great promise for the transmission of quantum information in NH systems.
\par Based on this discussion, one might expect that, similar to delta driving, step driving would produce a completely unidirectional skin effect. Our observations indicate that for certain intermediate values of $t_V$, such as $t_V = n\pi/2$, a bidirectional skin effect emerges. This bi-directionality is crucial, as it ensures at the vicinity of $t_V = n\pi$, the skin modes align at the same edge for each $n$. To understand this further, let us start with an infinitesimally small value of $t_V$, say $t_V = 0.001$, where, depending on the sign of $\gamma$, all the states are localized at one end of the system. For example, assume all the states are initially localized at the left end. As $t_V$ increases, some states begin to appear at the opposite end, that is the right end of the system. When $t_V$ reaches exactly $\pi/2$, the states are evenly distributed between both the edges of the ladder. As $t_V$ increases further, all the states gradually shift towards the right end until $t_V$ approaches $t_V = \pi$. Then, at exactly $t_V = \pi$, the system loses NHSE. However, once 
$t_V$ exceeds the value $t_V = \pi$, the NHSE reemerges, but with a reversal of the direction, causing all states to accumulate at the left end again, similar to what was observed at $t_V = 0.001$, and the process repeats for each cycle of $n\pi$ in $t_V$. Moreover, the bi-directionality emerges owing to the CS of the system. Consequently, at $t_V = \pi/2$, all the states with positive (negative) energies accumulate to the right (left) end leading to an equal probability of distribution across both the edges.
\begin{figure}[t]
    \begin{subfigure}[b]{\columnwidth}
         \includegraphics[width=0.7\columnwidth]{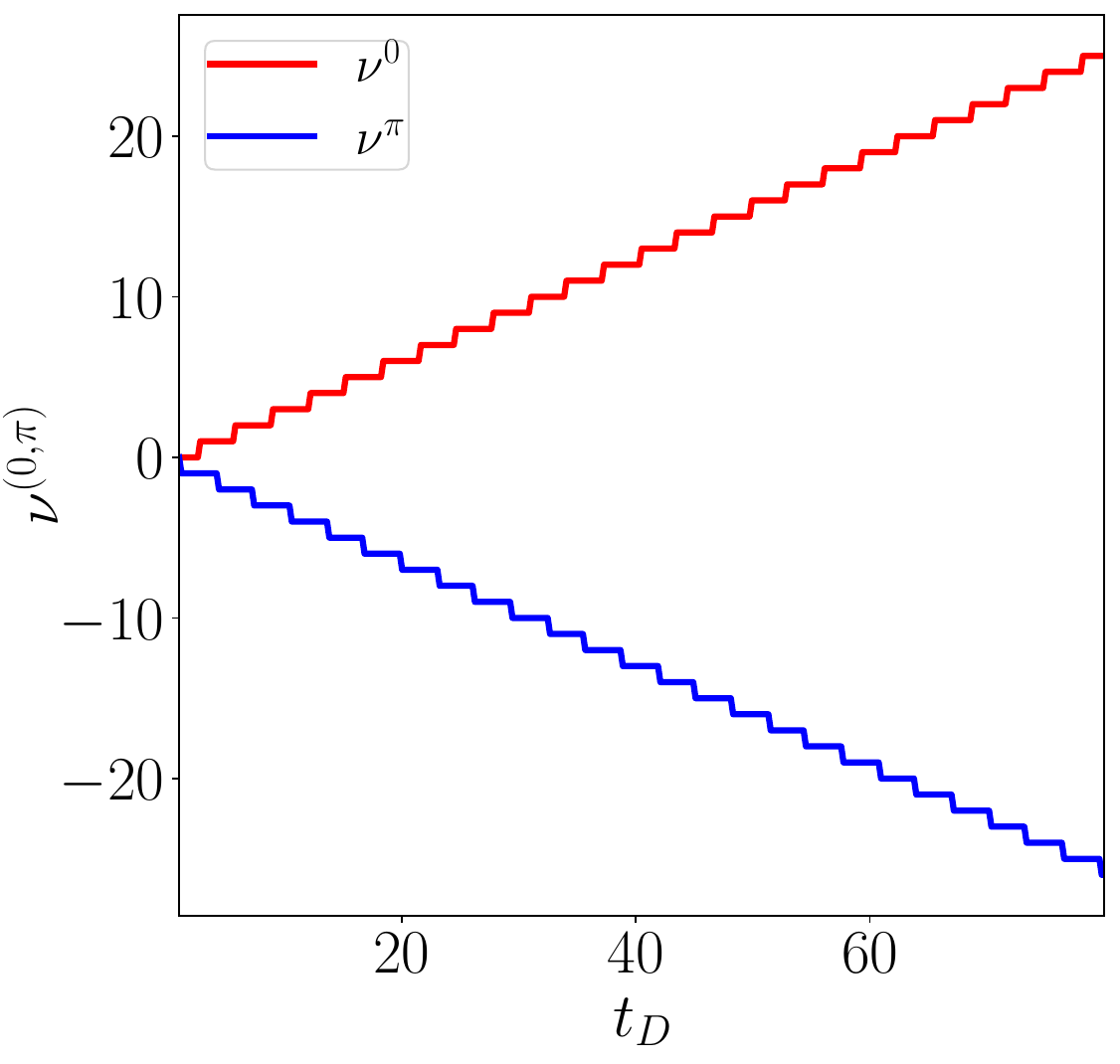}
         \label{1}
     \end{subfigure}
\caption{{Figure shows the step-like growth of winding numbers $\nu^{0,\pi}$ as a function of $t_D$ provided $t_V = n\pi$. The rest of the parameters are chosen as, $t_H = \pi/2$, $T =2$, and $\gamma=0.4$.}} 
\label{Figure_8}
\end{figure}
\par At this stage it is crucial to provide a fundamental explanation for these unique non-Hermitian features, which are otherwise absent in any type of drive. By closely examining the expressions for the \(d\)-vectors associated with the step drive model in both time frames, we observe that at the Floquet Hermitian points, the vertical hopping \(t_V\) effectively drops out of the expressions. Specifically, for \(t_V = n\pi\), \(d_{1,x}\) simplifies to, 
\begin{equation}
\begin{split}
    d_{1,x} &= \Big[\sin(t_V + 2t_D \cos(k))\Big] \times \Big[\cos(2t_H \sin(k) - i\gamma)\Big] \\ &= -\Big[\sin(2t_D \cos(k))\Big] \times \Big[\cos(2t_H \sin(k) - i\gamma)\Big],
    \end{split}
\end{equation} where the dependence on $t_V$ no longer exists. A similar scenario occurs for the other components of the \(d\)-vectors as well in each time frame. Hence, at the Floquet Hermitian points, the contribution from the vertical hopping amplitude gets eliminated. Now, if we recall, from the static perspective \cite{NHCreutz1}, the radius of the GBZ is typically determined by the ratio of upward to downward vertical hopping amplitudes, which can be written as \cite{NHCreutz1},
\begin{equation}
\beta = |r|e^{ik} = \sqrt{\left| \frac{t_V + \gamma}{t_V - \gamma} \right|} e^{ik}
\end{equation}
However, in the driven scenario, at the Floquet Hermitian points, the absence of the vertical hopping can lead to a radius equal to unity. This results in the GBZ assuming the form of a unit circle, which further implies the absence of skin effect. Nevertheless, the NHSE can re-emerge with slight deviations from the condition \(t_V = \pi\), where, depending on whether \(t_V\) exceeds or falls below \(\pi\), all the wavefunctions accumulate at one edge or the other. 
\par By utilizing the coordinates of the GBZ from both time frames, similar to what we have done for the delta drive, we can extract the non-Bloch invariants $(\nu^{0,\pi})$ from Eq. \ref{symmetric_frame}. Fig. \ref{Figure_6}b illustrates variation of $\nu^{0,\pi}$ as a function of $t_V$, which shows excellent agreement with the OBC spectrum depicted earlier in Fig. \ref{Figure_6}a. Furthermore, the step drive is expected to exhibit higher winding numbers in the low-frequency regime, similar to the delta drive. Particularly at the Floquet Hermitian points $(t_V = n\pi)$, intriguing features of the winding numbers arise that were not observed in the delta-driven case. For example, the dependence of $\nu^{0,\pi}$ on $t_D$ (provided $t_V = n\pi$) reveals a step-like growth in the winding numbers (Fig. \ref{Figure_8}), which may be considered analogous to the plateau behavior of Hall resistivity, as seen in the quantum Hall effect experiments.
\subsubsection{\textbf{Exceptional points}}
\begin{figure}[t]
\hspace{-1cm}
    \begin{subfigure}[b]{0.5\columnwidth}
         \includegraphics[width=1\columnwidth]{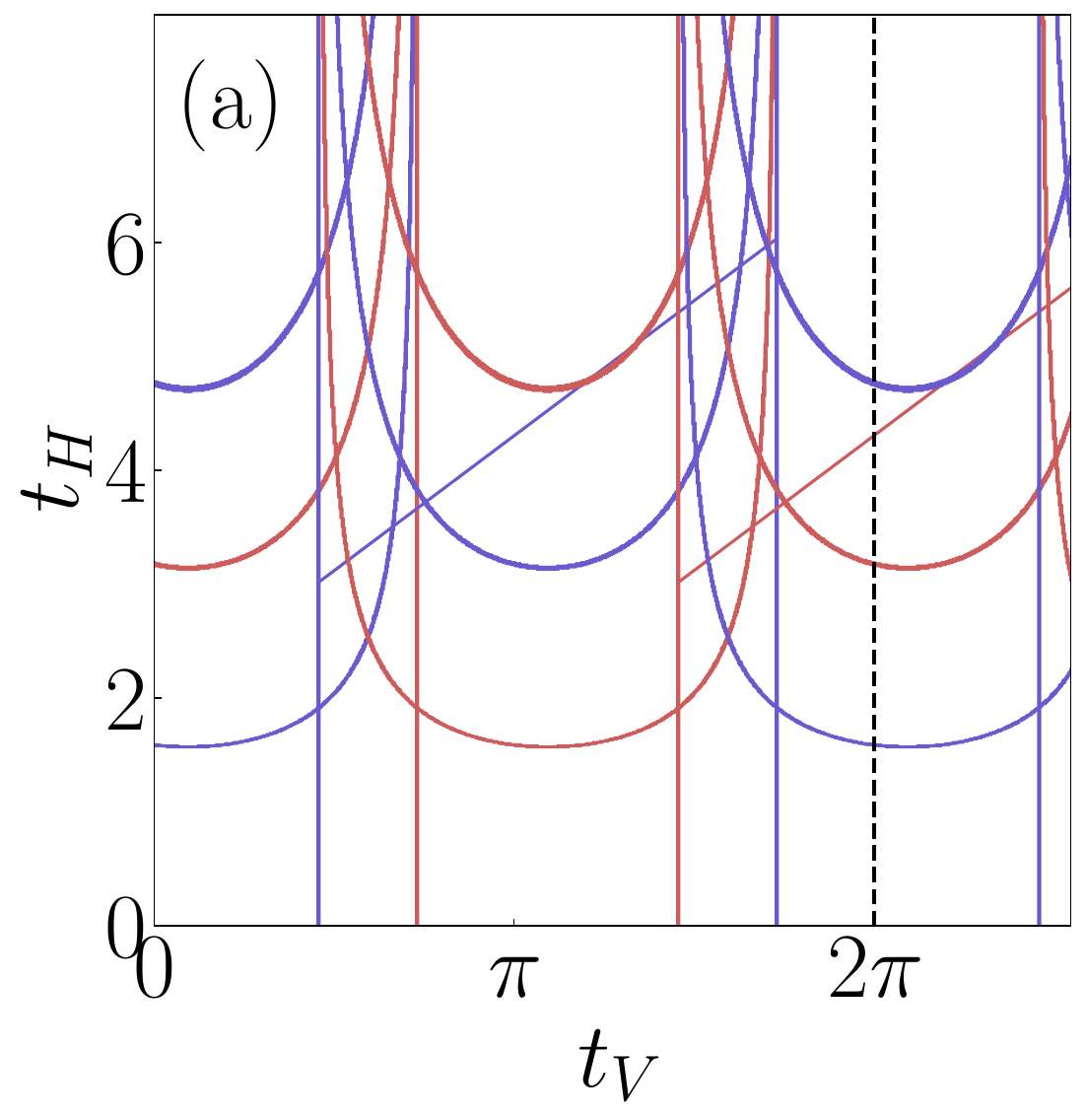}
         \label{Figure_9a}
     \end{subfigure}
     \begin{subfigure}[b]{0.5\columnwidth}
         \includegraphics[width=1.05\columnwidth]{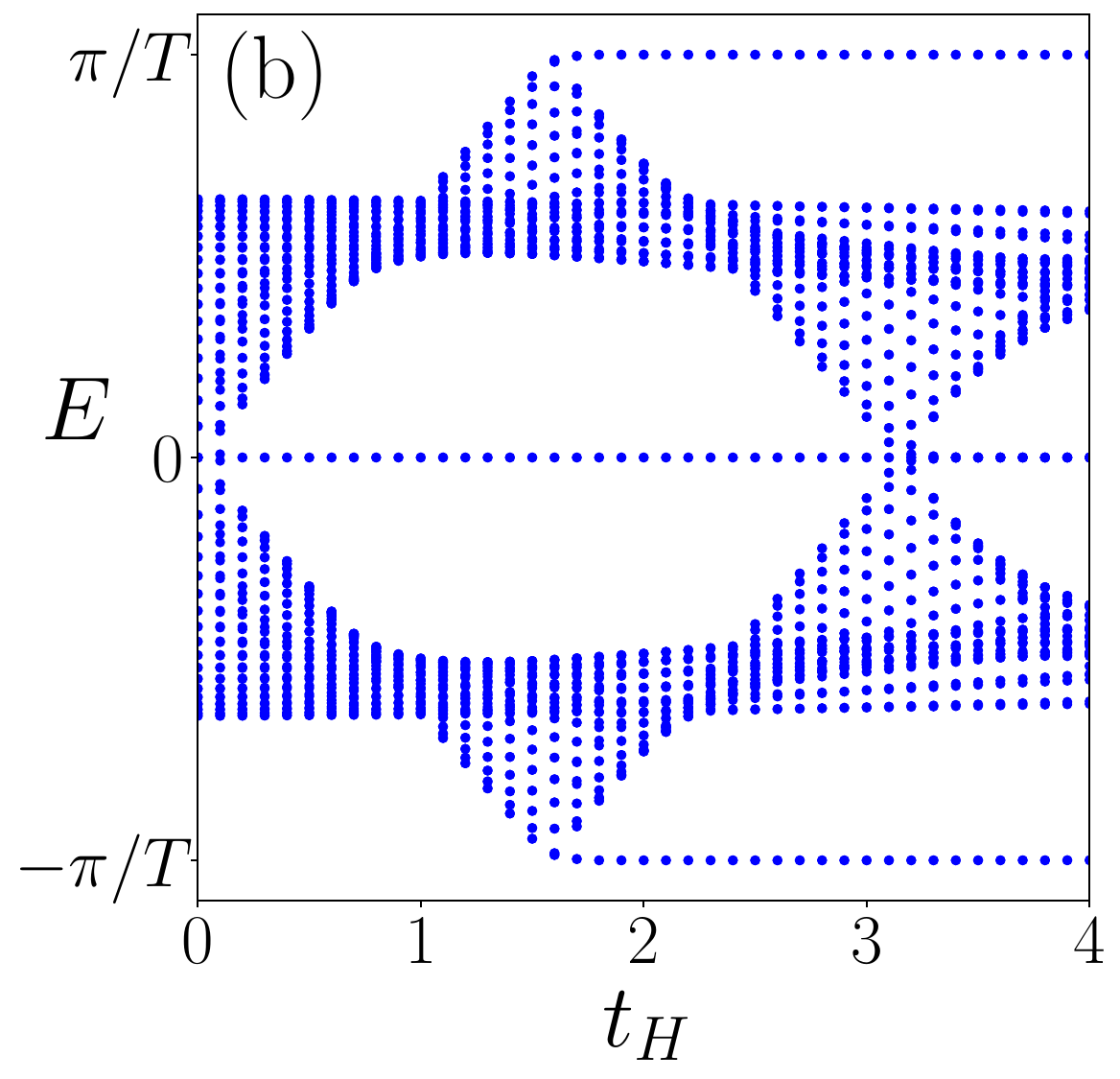}
         \label{Figure_9b}
     \end{subfigure}
\caption{{Panel (a) depicts the phase boundary diagram obtained using the trajectories of the EPs corresponding to the spectrum gap closing at quasienergy zero (red solid lines) and $\pi$ (blue solid lines). Panel (b) shows the quasienergy spectrum as a function of $t_H$ potted for $t_V = 2\pi$. The observed phase transitions, marked by gap closures, align precisely with the phase boundaries shown in panel (a). The rest of the parameters are chosen as, $T=2$, $\gamma=0.4$, and $t_D=1$ }} 
\label{Figure_9}
\end{figure}

The location of the EPs can be found by obtaining the expression of energy $E(k)$ for the step drive given as,
\begin{equation}
    \cos[E(k)] = \cos[d_x (k)] \cos [ d_z (k) ].
    \label{quasienergy_step_drive}
\end{equation}
See Appendix \ref{appendix} for a detailed derivation of $E(k)$. Thus, when the quasienergy gap closes at $E(k)=0(\pm \pi)$, we shall have $\cos[E(k)] = +1(-1)$ respectively. As a result, the location of the EPs can be identified by simultaneously satisfying the following conditions:
\begin{subequations}
\begin{align}
t_V + 2t_D \cos k &= m\pi \pm \arccos\left[\frac{1}{\cosh(\gamma)}\right], \\
2t_H \sin k &= n\pi.
\end{align}
\end{subequations}
where $m$ and $n$ are integers of the same (opposite) parities (even/odd) if the gap closes at $E(k)= 0 (\pm \pi)$. Subsequently, the loci of the EPs can be determined by varying the value of $ t_V $ to observe whether the gap closes at the center or the edge of the FBZ (as we did for Fig. \ref{Figure_5}). Furthermore, it is interesting to observe that by combining these gapless conditions and applying the trigonometric identity $\sin^{2}(k) + \cos^{2}(k) = 1$, one can ascertain the trajectory of EPs within the parameter space, as described by,
\begin{equation}
    \frac{1}{(2t_D)^2} \left( m\pi \pm \arccos\left[\frac{1}{\cosh(\gamma)}\right] - t_V \right)^2 + \frac{n^2 \pi^2}{(2t_H)^2} = 1
\end{equation}
These trajectories can form topological phase boundaries which ultimately separate different NH phases. For instance, in Fig. \ref{Figure_9}a, we have presented an example of such a phase boundary diagram obtained using the trajectories of EPs for the specific choice of parameters say, $t_D=1,T=2$. Here red (blue) lines represent different gap closings at quasienergy zero $(\pm \pi)$ respectively. Thus, for any non-zero $ \gamma $, it is reasonable to assume that the phase boundaries, or the gap-closing trajectories, can separate distinct NH phases which are shifted and deformed from their Hermitian counterparts by approximately $( \pm \arccos{[\frac{1}{\cosh(\gamma)}]} $). Therefore, the different band touching conditions can not be altered by tuning parameters other than $\gamma$, indicating the genuine robustness of the NH features of the system.
\par Additionally, since the trajectories of EPs were derived using the conventional BZ, the resulting phase boundary diagram is not expected to agree with the OBC spectrum. However, when we select a specific parameter, namely $t_V = 2\pi$, where the skin effect is absent, it becomes feasible for the phase boundary diagram to indicate phase transitions similar to that of the OBC spectrum. This is corroborated by our observations along the vertical line at $t_V = 2\pi$ in the phase boundary diagram (Fig. \ref{Figure_9}a), where we consistently find that the appearance of red (blue) lines corresponds to gap closings at quasienergies zero ($\pi$) in the OBC spectrum, (shown in Fig. \ref{Figure_9}b). Moreover, the value of $\nu^{0,\pi}$ can show a quantized jump every time when the trajectory of the gap closure at quasienergy zero ($\pm \pi$) is crossed in the parameter space. These observations deduce that the trajectories of gap closing at $E=0, \pm \pi$ are indeed the phase boundaries for different
NH FTPs, each being characterized by a pair of winding numbers, $\nu^{0}$ and $\nu^{\pi}$.
\begin{figure}[t]
    \begin{subfigure}[b]{\columnwidth}
         \includegraphics[width=1\columnwidth]{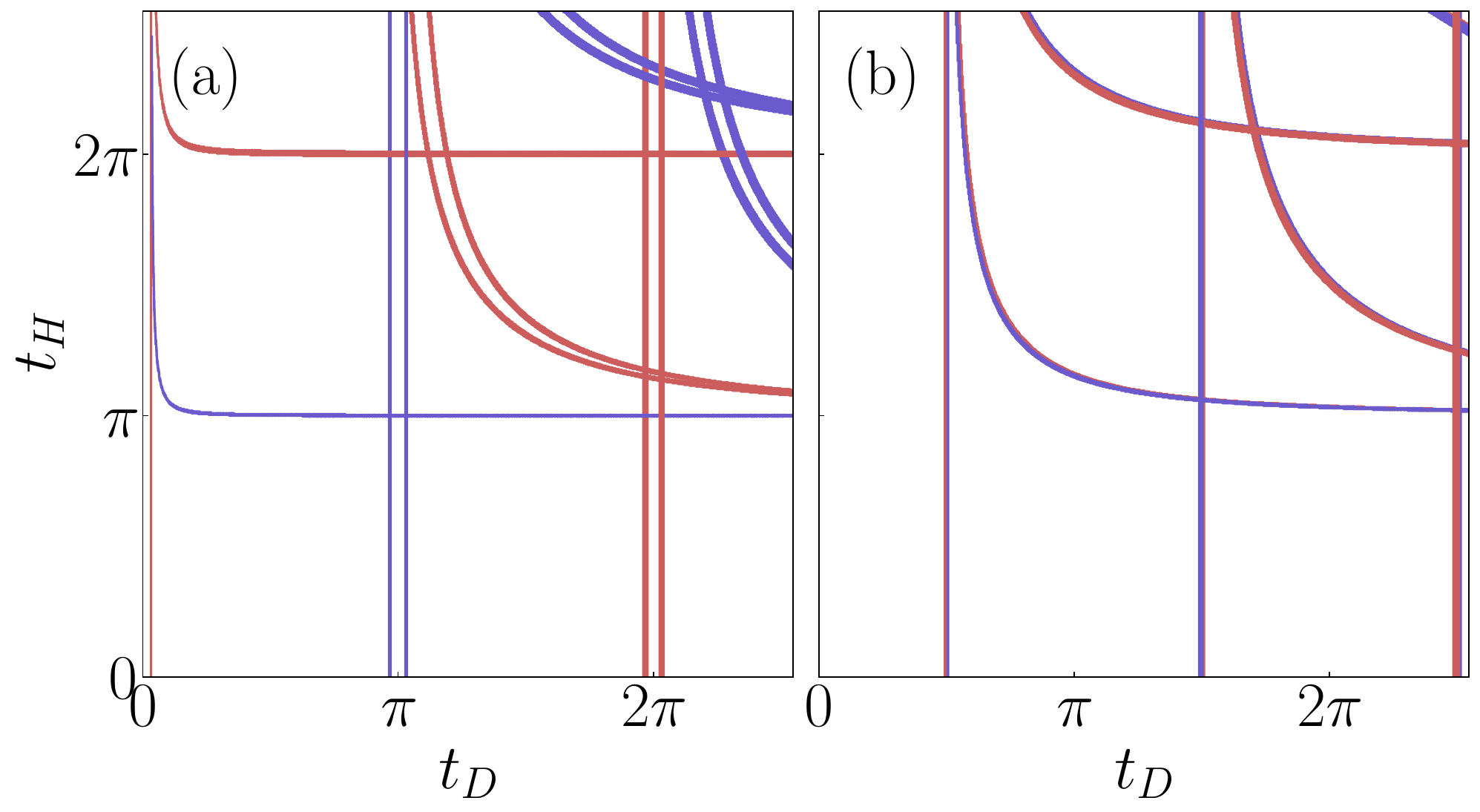}
         \label{1}
     \end{subfigure}
\caption{{The figure depicts the phase boundary diagram obtained using the trajectories of the EPs corresponding to the gap closing of the quasienergy spectrum at zero (red solid lines) and $\pi$ (blue solid
lines) for $\gamma = 0.05$ (panel a) and $\gamma=5$ (panel b) respectively, with the condition $t_V=n\pi$.}} 
\label{Figure_10}
\end{figure}
\par In addition, we have provided two additional examples in the phase boundary diagram, one where $\gamma$ is very small (close to the Hermitian limit) and another where $\gamma$ is significantly large. Fig. \ref{Figure_10}a denotes the phase boundary diagram with a very small NH coupling say $\gamma=0.05$. Similar to the behavior observed in the delta drive, each phase boundary corresponding to the Hermitian limit now splits into a pair of closely spaced trajectories formed by EPs. On the other hand, for a relatively large NH coupling, such as $\gamma = 5$ (provided $(t_V = n\pi$)), intriguing features arise that were absent in the delta drive. For instance, the phase boundaries under such enormous non-Hermiticity get shifted in such a way that the trajectories for the gap closure at zero quasienergy nearly coincide with those at the quasienergy $\pi$ (Fig.~\ref{Figure_10}b). This indicates that crossing these phase boundaries will result in phase transitions where the gap closure at both zero and $\pi$ quasienergy occur simultaneously. Such an intriguing feature is entirely lacking from its Hermitian as well as delta-driven counterpart. A recent investigation \cite{period2t1,period2t2} demonstrated that such simultaneous superposition of zero and $\pi$ edge states can lead to a symmetry-protected discrete time crystal phase, referred to as `period-2$T$ Floquet time crystal phase', offering new possibilities for non-Abelian braiding and quantum computation in Floquet topological systems.
\begin{figure*}[t] 
\begin{minipage}{\linewidth}
        \begin{subfigure}[t]{0.95\columnwidth}
            \includegraphics[width=\linewidth]{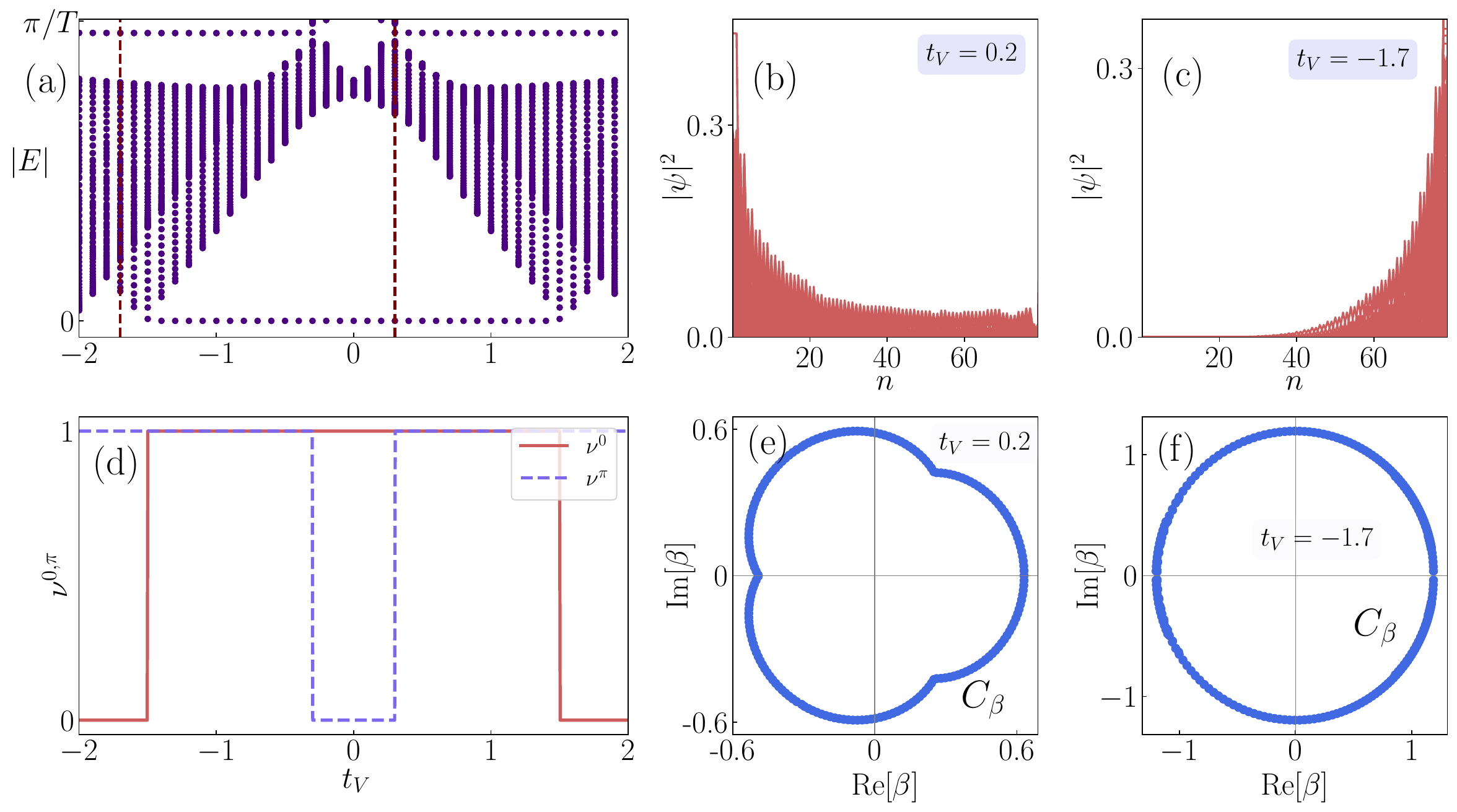}
        \end{subfigure}%
\caption{Panel (a) shows the Floquet quasi-energy spectrum corresponding to the harmonic drive, plotted as a function of the $t_V$. Panels (b) and (c) describe the probability distribution of the eigenstates, and their corresponding GBZ ($C_\beta$) have been shown in panels (e) and (f). The coordinates of GBZ have been utilized to compute $\nu^{0,\pi}$ as shown in panel (d) which correctly coincides with panel (a). The rest of the parameters are chosen as, $t_H=0.75$, $t_D=0.75$, $V_0=0.5$, $\gamma=0.4$, and $\omega=3.5$.}
\label{Figure_11}
    \end{minipage}
\end{figure*}
\subsection{\label{sec:level3c}Harmonic drive}
Unlike the step or the delta drive, the Floquet operator for a harmonic drive cannot be expressed as a product of two exponential operators. Instead, it must be calculated by breaking the entire period into numerous small steps and performing time-ordered multiplication. This makes deriving the exact expression for the components of the $d$-vectors analytically challenging, which in turn complicates solving the characteristic polynomial, $\det[H(\beta) - E] = 0$. To address this issue, one can map the entire problem into a rotating frame approximation followed by a Magnus expansion \cite{magnus1,magnus2,magnus3} which will effectively yield exact expressions for the $d$-vectors capable of reproducing the results from the OBC spectrum provided the frequency is large. 
\par To begin with, let us consider a harmonic drive associated with the vertical hopping given as,
\begin{equation}
    H_{V}(t)=(2V_{0}\cos{\omega t} + t_V)\sum_{n}(a_n^{\dagger} b_n+a_n b_n^{\dagger}).
\end{equation}
The corresponding quasienergy spectrum has been plotted in Fig. \ref{Figure_11}a, from which we have picked two specific values of $t_V$, namely $t_V = 0.2$ and $t_V = -1.7$, to verify the NHSE by showing the accumulation of wavefunctions in Figs. \ref{Figure_11}b and \ref{Figure_11}c. Further, to analyze the correct analytical behavior of the $d$-vectors associated with the Floquet Hamiltonian, we resort to the high-frequency approximation, which involves a rotating frame transformation in the Floquet formalism. In the rotating frame, given by a unitary transformation $S(t)$, the transformed Floquet Hamiltonian takes the form,
\begin{equation}
\begin{split}
\tilde{H}_{k}(t)   = S^{\dagger}(t)H_k(t)S(t)-iS^{\dagger}(t)\dot{S}(t).
\end{split}
\end{equation}
We may choose to work with some particular choices for rotating frames, where the unitary transformation is defined as,
\begin{equation}
    S(t)= \text{exp}[-i \hat{d}_{\beta} \cdot \vec{\sigma} \omega t /2],
\end{equation}
where, $\hat{d}_{\beta}= \mathbf{d}_{\beta}/|d_{\beta}|$, and $|d_{\beta}|= E_0(\beta) = \sqrt{d^2_x(\beta) + d^2_z(\beta) }$. Moreover, for the simplicity of calculation, we aim to work on the frequency domain so that it allows us to write the effective Hamiltonian as an expansion in powers of the inverse of frequency, also referred to as the  Magnus expansion. This yields a Floquet effective Hamiltonian given by,
\begin{equation}
\begin{split}
    H_{\text{eff}} & = H_{0}(\beta) + \sum_{m\neq 0}\frac{[H_0,H_m]}{m \omega} + \sum_{m > 0}\frac{[H_m,H_{-m}]}{m \omega}, 
\end{split}
\label{magnus}
\end{equation}
where,
\begin{equation}
    H_{\pm m} = \frac{1}{T}\int^T_0 {H(t)e^{\pm i m \omega t} dt}.
    \label{Floquet_replica}
\end{equation}
Note that, owing to the mathematical form of the drive the only surviving terms in Eq. \ref{Floquet_replica} are those with $m=0$ and $\pm 1$. As a result, the infinite series in Eq. \ref{magnus} can be truncated at the first-order limit, leading to an effective Hamiltonian of the form given as,
\begin{equation}
    H_{\text{eff}} = d_x^{\prime} (\beta) + d_z^{\prime} (\beta),
\end{equation}
where,
\begin{subequations}
    \begin{align}
        d_x^{\prime}(\beta) &= \left[ 1 - \frac{\omega}{2 |d_{\beta}|} \right] d_x(\beta) + \frac{V_0}{2|d_{\beta}|^2} d^2_z(\beta), \\
        d_z^{\prime}(\beta) &= \left[ 1 - \frac{\omega}{2 |d_{\beta}|} \right] d_z(\beta) + \frac{V_0}{2|d_{\beta}|^2} d_x (\beta) d_z (\beta).
    \end{align}
    \label{d_vectors_harmonic}
\end{subequations}
Interestingly, for small driving strengths, the last terms in Eq. \ref{d_vectors_harmonic} can be disregarded. This simplifies the characteristic polynomial for $\beta$ ($\det[H(\beta)-E]=0$) to a second-order form. Hence, it is plausible that for weak driving strengths, the GBZ may assume a circular shape with a uniform radius as the static case.
\par In Figs. \ref{Figure_11}e and \ref{Figure_11}f, we present the plot of $C_{\beta}$ corresponding to the values marked in the quasienergy spectrum shown in Fig. \ref{Figure_11}a, based on the specific choices, namely, $t_H=0.75$,  $t_D=0.75$, $V_0 = 0.5$, $\omega=3.5$ and $\gamma=0.4$. Similar to the delta-driven scenario, where all energy states are localized at the left (or right) end (Figs. \ref{Figure_11}b and \ref{Figure_11}c), the corresponding GBZ has a radius lesser (greater) than 1.
\par Further, the coordinates of the GBZ, derived from solving the characteristic polynomial in terms of the $d$-vectors, will be employed in both time frames to determine the pair of the topological invariants $(\nu^{0,\pi}$). However, the derivation of the time frame for the harmonic drive differs significantly from the approaches we used earlier for the delta and the step drives. Ensuring the necessary symmetries of the system, there has to be an intermediate time, $t=T/2$ that splits the period into two symmetrical parts \cite{asbothwinding1,asbothwinding2}. Let $F$ and $G$ denote the time evolution of the first and second part of the cycle respectively, that is,
\begin{equation}
    F = \mathcal{T} e^{-i \int_{0}^{t} H(t) dt} ~ ; ~ G = \mathcal{T} e^{-i \int_{t}^{ T} H(t) dt}. 
\end{equation}
Now, if these two operators are chiral symmetric partner similar to Eq.~\ref{chiral_partner}, then one can form the two symmetric frames as, $\hat{U}_1 = \hat{F} \hat{G}$ and $\hat{U}_2 = \hat{G} \hat{F}$. Building on these two time frames, each of the non-trivial phases can be characterized via a pair of invariant similar to Eq.~\ref{symmetric_frame}.
\par In Fig. \ref{Figure_11}c, we have shown the variation of $\nu^{0,\pi}$ as a function of $t_V$ which clearly agrees with the OBC spectrum. It is noteworthy, that unlike the delta or the step drive, the harmonic drive can not host higher winding numbers. This is due to the fact that the computation of the winding numbers are based on the $d$-vectors obtained from a high-frequency expansion, and it is expected that at high-frequency regime the system can not have multiple gap closing, and hence multiple edge modes. Nonetheless, the approach we have discussed here can very much be utilized over a range of frequencies, revealing intriguing NH features owing to its periodic modulation. Furthermore, we note that certain points in the quasi-energy spectrum (Fig. \ref{Figure_11}a) appear at \( |E| > \pi/T \) which may be understood as follows. This spectrum is obtained via the Shirley-Floquet formalism \cite{floquetformalism3,SF2,SF3} and comprises of three replicas of the static branches, each separated from the other by an integer multiple of the driving frequency (\( m\omega \)). Consequently, the points with \( |E| > \pi/T \) in Fig. \ref{Figure_11}a emerge from the bulk states of the upper replica (\( m = +1 \)), which are intrinsically linked to the central replica (\( m = 0 \)).
\par Moreover, it is worthwhile to mention that both harmonic and delta drive are expected to yield similar results. This is due to the fact that a delta drive can effectively be viewed as a mutual superposition
of multiple harmonic drives with appropriately tuned frequencies. However, the methodologies for computing the
GBZ in these two cases are entirely different which we have comprehensively discussed so far. Consequently, EPs are supposed to bear close quantitative resemblance to those obtained for the delta drive, albeit their precise enumeration are significantly more complicated due to the complex structure of the $d$-vectors (Eq. \ref{d_vectors_harmonic}).
In conclusion, our formalism creates new opportunities for uncovering non-trivial NH characteristics through the modulation of various driving protocols, offering exciting prospects for exploring NH effects in driven systems. Moreover, each driving generates unique features unattainable in static or Hermitian systems, as we have summarized in Table \ref{Table}.
\section{\label{sec:level4}conclusion}
In this work, we explore the interplay between periodic driving and non-Hermitian effects in a quasi-1D ladder system. This inherent quasi-1D structure, in presence of an on-site complex potential effectively creates a non-reciprocal hopping, which eventually yields an NHSE. We have adopt a generalized Floquet non-Bloch formalism capable of recovering the BBC even under periodic modulation. To extend this formalism to various driving protocols, we analyze a delta drive, a step drive, and a harmonic drive, each yielding unique features unattainable in the static NH cases. For example, while both the delta and the harmonic drives exhibit a unidirectional skin effect, the step drive manifests both unidirectional and bidirectional skin effects. Interestingly, although the Hamiltonian at different quenches under the step drive may not exhibit NHSE, yet the effective Hamiltonian over a complete period demonstrates NHSE. Moreover, specific points within the parameter space of the step drive model are found to be free from skin effect, which we designate as Floquet Hermitian points. 
\begin{table}[t]
    \centering
    \renewcommand{\arraystretch}{1.2} 
    \setlength{\tabcolsep}{1.6pt} 
    \footnotesize 
    \begin{tabular}{|p{2.15cm}|p{1.9cm}|p{1.9cm}|p{1.9cm}|}
        \hline
        \textbf{} & \textbf{Delta Drive} & \textbf{Step Drive} & \textbf{Harmonic Drive} \\
        \hline
        NHSE & Unidirectional & Bidirectional/ Unidirectional & Unidirectional \\
        \hline
        Floquet Hermitian Point & \quad Absent & \quad Present & \quad Absent \\
        \hline
        Higher Winding Number & \quad Possible & \quad Possible & ~Not Possible \\
        \hline
        Trajectories of EPs for Zero and $\pi$ modes & Never coincide & ~Can coincide & Never coincide \\
        \hline
    \end{tabular}
    \caption{Table depicting the comparison of distinct non-Hermitian features corresponding to different driving protocols. Here, Floquet Hermitian points refer to points in the parameter space for the non-Hermitian scenario where the system can be free from skin effect under the application of periodic driving.}
    \label{Table}
\end{table}
\par While the skin effect in all the cases disrupts the BBC, construction of a GBZ using a non-Floquet Bloch formalism could pave the way towards a Floquet non-Bloch invariant. However, it is evident that periodic driving breaks chiral symmetry, leading us to employ a symmetric time frame approach to construct the GBZ for both the delta and the step drives. Whereas, a high-frequency expansion for the harmonic drive is employed which allows us to derive exact expressions for the $d$-vectors, thereby contributing to the construction of the GBZ. 
\par Due to the time periodicity, we also observe longer-range interactions in the low-frequency regime, which amplify the non-reciprocal effects induced by the non-Hermiticity term. This amplification causes the skin modes to localize at one end of the system, filling either the entire upper or the lower diagonal of the Hamiltonian matrix, however, leaving the rest completely empty. Furthermore, our investigation of the EPs associated with each of the driving schemes reveals that each pair of the EPs emerges from the degenerate points of their Hermitian counterparts. Notably, in the strong NH limit for the step drive, the trajectories of the EPs diverge such that each phase transition now coincides with simultaneous zero and $\pi$ quasienergy gap closings.
\section*{Acknowledgements}
KR and KG sincerely acknowledge Dipendu Halder, Latu Kalita, and Sankalan Jain for fruitful discussions.
\appendix
\section{\label{appendix1}Symmetries associated with the NH version of Creutz ladder}
\begin{table*}[t]
\centering
\setlength{\tabcolsep}{16pt} 
\renewcommand{\arraystretch}{1.5} 
\begin{tabular}{l l l}
\hline\hline
Symmetry & Equation & Operator \\ \hline
TRS, $T_{+}T_{+}^* = \pm 1$ & $H(-k) = T_{+}H^*(k)T_{+}^{-1}$ & $\sigma_x$ \\
PHS, $C_-C_-^* = \pm 1$ & $H(-k) = - C_-H^T(k)C_-^{-1}$ & $\times$ \\
TRS$^\dagger$, $C_+C_+^* = \pm 1$ & $H(-k) = C_+H^T(k)C_+^{-1}$ & $\times$ \\
PHS$^\dagger$, $T_-T_-^* = \pm 1$ & $H(-k) = -T_-H^*(k)T_-^{-1}$ & $\sigma_z$ \\
CS, $\Gamma^2 = 1$ & $H(k) = -\Gamma H^{\dagger}(k)\Gamma^{-1}$ & $\times$ \\
SLS, $S^2=1$ & $H(k) = -S H(k)S^{-1}$ & $\sigma_y$ \\ \hline\hline
\end{tabular}
\caption{Symmetries and their corresponding notations for the non-Hermitian Creutz ladder. Each symmetry has its own constraint equation, which is satisfied by the symmetry operator acting on the Hamiltonian. The symbol `$\times$' denotes that, under no condition, can the non-Hermitian Creutz ladder preserve the corresponding symmetry.}
\label{Table2}
\end{table*}
While the Altland–Zirnbauer (AZ) classification \cite{10fold} is well-established for Hermitian systems where symmetry classes are determined by the presence (or absence) of time-reversal (TRS), particle-hole (PHS), and chiral symmetries (CS), non-Hermitian systems introduce significant modifications to these classifications \cite{NH10fold}. The fundamental distinction lies in the fact that, for non-Hermitian systems, complex conjugation and transposition are no longer equivalent operations, that is, $H^* \neq H^T$. This leads to a ramification of PHS, where the conditions, \(CH^*C^{-1} = -H\) and \(CH^T C^{-1} = -H\) emerge as distinct cases. Moreover, TRS, being an antiunitary symmetry, follows either of the conditions, \(T_+H^*T_+^{-1} = H\) or \(T_-H^*T_-^{-1} = -H\) in the Hermitian limit. This framework allows for a direct mapping of a non-Hermitian Hamiltonian \(H\) to \(iH\) with a one-to-one correspondence. Hence, \(T_+H^*T_+^{-1} = H\) implies \(T_+(iH)^*T_+^{-1} = -iH\), thereby establishing \(T_+ \equiv T_-\) and unifying these two symmetries. Interestingly, it is noteworthy that \(T_-\) can be interpreted as a variant of the PHS, which we denote as PHS\(^{\dagger}\). Similarly, a variant of the TRS, denoted by TRS\(^{\dagger}\), can be defined by applying \(C\) to the transpose of the Hamiltonian, expressed as \(C_-H^T C_-^{-1} = H\).
\par Further, in context of the non-Hermitian symmetry classification, the roles of the chiral and sublattice symmetries deviate significantly from those in Hermitian systems. While in the Hermitian case the product of TRS and PHS results in CS which ultimately coincides with the sublattice symmetry (SLS), this equivalence does not hold for non-Hermitian systems. Instead, SLS emerges as a distinct internal symmetry, described by \(S H S^{-1} = -H\), where \(S\) is a unitary operator satisfying \(S S^{\dagger} = 1\). This distinction results from the ramification and unification of symmetries, expanding the 10-fold AZ classification for Hermitian systems to a 38-fold classification for non-Hermitian systems \cite{NH10fold}, which can extend further to 54 symmetry classes under the application of periodic driving \cite{NH10foldfloquet}. Moreover, depending upon whether the sublattice symmetry commutes or anticommutes with the TRS and PHS, one can assign distinct invariants for the system. Therefore, systems with a point-gap spectrum, essential for the emergence of the non-Hermitian skin effect, can be classified to \(S_-\) which is characterized by a $\mathbb{Z}$ invariant \cite{NH10fold}, which in this case is the winding number. To define such an invariant, the Hamiltonian is required to be expressed in the sublattice basis in an off-diagonal form, given as 
\begin{equation}
    H = \begin{pmatrix} 0 & H_+ \\ H_- & 0 \end{pmatrix},
\end{equation} where the \(+\) and \(-\) signs depend on the signs associated with the non-Hermitian terms (that is $+\gamma$ or $-\gamma$). Moreover, in the Hermitian limit, \( H_+ = H_-^{\dagger} \), recovering the matrix representation in the chiral basis, thereby unifying the chiral and the sublattice symmetries. However, periodic driving breaks the sublattice (or chiral in the Hermitian limit) symmetry, necessitating the usage of symmetric time frames, as elaborated in the main text of the manuscript. Moreover, the explicit forms and the fundamental definitions for each of the symmetry operations associated with our non-Hermitian system have been illustrated in Table \ref{Table2}.
\section{\label{appendix}Evaluation of $\mathbf{\textit{d}}$ -vectors and expression for quasienergies in two symmetric time frames}
Let us consider a generic expression of the Floquet evolution operator which can be written as a product of two exponential matrices, given as
\begin{equation}
   \hat{U}(T) = e^{-iT_m \mathbf{h}_m(k) \cdot \boldsymbol{\sigma}} \quad (m = 1,2).
   \label{A1}
\end{equation}
Now, according to Euler's identity of the Pauli matrices,
\begin{equation}
    e^{-iT_m \mathbf{h}_m(k) \cdot \boldsymbol{\sigma}} = \cos{|T_m \mathbf{h}_m|} \sigma_0 - i \sin{|T_m \mathbf{h}_m|} \hat{h}_m \cdot \boldsymbol{\sigma},
    \label{A2}
\end{equation}
where $\sigma_0$ is the identity matrix. The Floquet evolution operator can therefore be expanded using Eq. \ref{A2} as,
\begin{equation}
\begin{split}
    \hat{U}(T) & = e^{-iT_2 \mathbf{h}_2(k) \cdot \bm{\sigma}} e^{-iT_1 \mathbf{h}_1(k) \cdot \bm{\sigma}} 
\\ & = ( \cos|T_1 \mathbf{h}_1| \cos|T_2 \mathbf{h}_2| \\ &- \sin |T_1 \mathbf{h}_1| \sin |T_2 \mathbf{h}_2| \cos \alpha ) \sigma_0
\\ & - i \cos|T_1 \mathbf{h}_1| \sin |T_2 \mathbf{h}_2| \hat{h}_2 \cdot \bm{\sigma} 
\\ &- i \cos|T_2 \mathbf{h}_2| \sin |T_1 \mathbf{h}_1| \hat{h}_1 \cdot \bm{\sigma}
\\ & + i \sin |T_1 \mathbf{h}_1| \sin |T_2 \mathbf{h}_2| ( \hat{h}_1 \times \hat{h}_2 ) \cdot \bm{\sigma},
\end{split}
\label{A3}
\end{equation}
where $\cos{\alpha} = \hat{h}_1 \cdot \hat{h}_2$. Furthermore, any Floquet evolution operator can be written in terms of the Floquet effective Hamiltonian as,
\begin{equation}
    \hat{U}(T) = e^{-i H_{\text{eff}} T}.
    \label{A4}
\end{equation}
Applying, the Euler's identity to Eq. \ref{A4}, one can get,
\begin{equation}
    e^{-i H_{\text{eff}} T} = \cos{[E(k)]}\sigma_0 - i \sin{[E(k)]} \mathbf{r} \cdot \bm{\sigma}.
    \label{A5}
\end{equation}
Comparing Eq. \ref{A3} with Eq. \ref{A5}, we can have the expression of quasienergies given as,
\begin{equation}
\begin{split}
    \cos{[E(k)]} = & \cos|T_1 \mathbf{h}_1| \cos|T_2 \mathbf{h}_2| \\ &- \sin |T_1 \mathbf{h}_1| \sin |T_2 \mathbf{h}_2| \cos \alpha,
\end{split}
\label{A6}
\end{equation}
while the exact expression of the $d$-vectors can be determined by extracting $\mathbf{r} \cdot \bm{\sigma}$ from Eq. \ref{A5}. Moreover, the quasienergy bands touch at both zero and $\pm \pi/T$, which occur when $E=1$ and -1 respectively. Therefore, the location of the EPs can be obtained from Eq. \ref{A6}, when it satisfies,
\begin{equation}
    T_m E_m = n \pi, \quad n \in \mathbb{Z},
\end{equation}
or
\begin{equation}
\begin{cases}
    \hat{h}_1 \cdot \hat{h}_2 = \pm 1, \\
    T_1 E_1 \pm T_2 E_2 = n \pi, \quad n \in \mathbb{Z}.
\end{cases}
\end{equation}
Now, building upon this formalism with two symmetric time frames, where the Floquet evolution operator is represented as a product of three exponential matrices, the expression for $\hat{U}(T)$ can be derived in a manner analogous to Eq. \ref{A3}, and can be represented as,
\begin{equation}
    \hat{U}_{j,k} = n_0 + i \mathbf{n_j}(k) \cdot \bm{\sigma} \quad ; (j = 1,2),
\end{equation}
where $n_0= \cos{[E(k)]}$, and
\begin{equation}
\begin{split}
    \mathbf{n_j} = & [\cos(|\mathbf{h}_j| T_j ) \sin(|\mathbf{h}_i| T_i ) (\hat{h}_j \cdot \hat{h}_i)
   \\ & + \sin(|\mathbf{h}_j| T_j ) \cos(|\mathbf{h}_i| T_i )] \hat{h}_j \\
    & + \sin(|\mathbf{h}_i| T_i ) [\hat{h}_i - (\hat{h}_j \cdot \hat{h}_i) \hat{h}_j],
    \label{A10}
\end{split}
\end{equation}
where $i,j$ denote the indices of corresponding to the two symmetric time frames, that is, either $i=1$, $j=2$, or vice versa. Therefore, corresponding to the step drive, by comparing the Floquet evolution operator (Eq. \ref{evolution_step_drive}) with Eq. \ref{A3}, we can assign $h_1 = d_z$, $h_2 = d_x$, and $T_1 = T_2 = T/2$. Finally, substituting these into Eq. \ref{A10} yields the expressions for the $d$-vectors in both the time frames, (presented in Eq. \ref{d_vectors_step_drive} of the main text). Similarly, the quasienergies of the Floquet effective Hamiltonian (Eq. \ref{quasienergy_step_drive}) are derived using Eq. \ref{A6}. In contrast, for the delta drive, we assign $h_1 = H_0$, $h_2 = V_0$, with $T_1 = T$ and $T_2 = 1$. Similarly, plugging these into Eq. \ref{A10} results in expressions for the components of the $d$-vectors, as presented in Eqs. \ref{d_vectors_delta_drive_1} and \ref{d_vectors_delta_drive_2}. It is worth noting that in the original time frame ($t = 0 \rightarrow T$), expanding the evolution operator $\hat{U}(T)$ using Eq. \ref{A3} reveals that the Floquet effective Hamiltonian can exhibit all the three components of the $d$-vector, as described by Eq. \ref{all_d_componets}. This expansion highlights the way in which the periodic drive disrupts CS.
\section{\label{appendix3}Numerical justification behind the truncation of the characteristic polynomial}
\begin{figure}[t]
\hspace{-0.9cm}
    \begin{subfigure}[b]{\columnwidth}
         \includegraphics[width=\columnwidth]{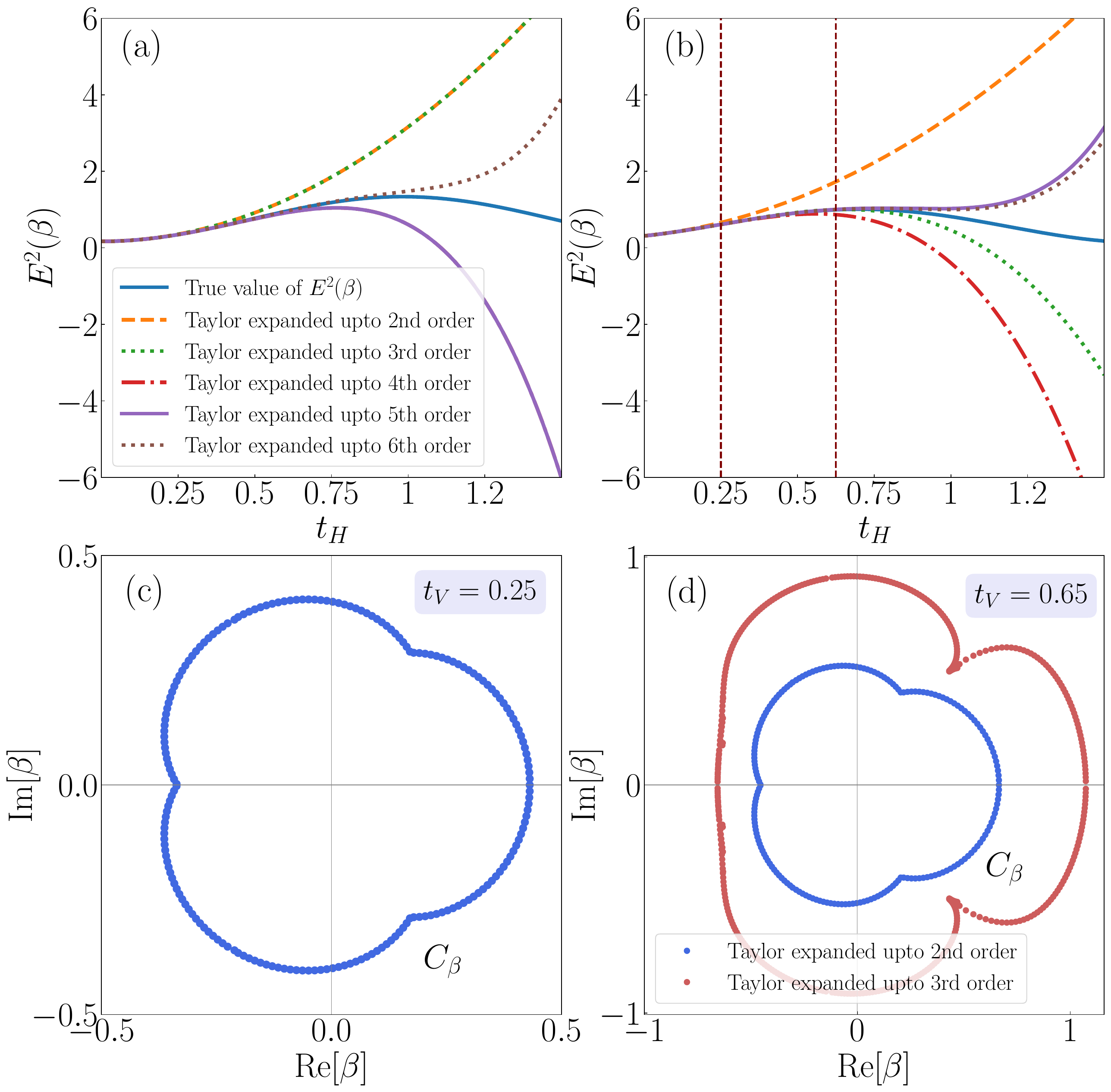}
         \label{1}
     \end{subfigure}
\caption{Panels (a) and (b) illustrate the variation of \( E^2(\beta) \) with respect to \( t_H \), alongside the Taylor-expanded functions, where different colors represent varying orders of truncation. In panel (a), the parameters are set to \( t_V = \pi \), \( \beta = 1 \), while in panel (b), the parameters are \( t_V = 3.5 \), \( \beta = 0.75 \) to distinguish the scenarios without and with skin effect. Panel (c) depicts the curvature of the GBZ for \( t_H = 0.25 \), using the same parameter values as panel (b). Notably, in this regime, the GBZ derived from different truncation orders converges perfectly. Conversely, panel (d) showcases the curvature of the GBZ for \( t_H = 0.65 \) with the parameters identical to those in panel (b). Here, a significant deviation emerges between the GBZ obtained from the second-order Taylor expansion and that from the third-order expansion.}
\label{Figure_12}
\end{figure}
As we have completed the step-wise procedure for evaluating $\beta$ (as discussed in Sec. \ref{GBZ}), it is now crucial to address the technical limitations and approximations inherent in our approach, particularly the justification for the order of truncation in the Taylor expansion. To this end, we have analyzed the behavior of the step drive model by plotting the function itself (as shown on the left-hand side of Eq. \ref{Taylor}) against $t_H$, with various orders of Taylor expansion depicted with different colors (see Fig. \ref{Figure_12}a). For this analysis, we have set $t_V = \pi$, a benchmark value which shows no skin effect, ensuring that $\beta=1$ remains valid for all the $t_H$ values. This setup provides a baseline for evaluating the accuracy of the truncated expansions. However, when $t_V \neq \pi$, the skin effect re-appears, causing $\beta$ to deviate from unity and become highly dependent on $t_H$. This complexity further gets intensified with the inclusion of higher-order truncations, making it challenging to assign consistent $\beta$ values across the full range of $t_H$. Therefore, focusing on the scenario of $t_V = \pi$ with $\beta=1$ provides a clearer picture for assessing the accuracy of different truncation orders. Now from Fig. \ref{Figure_12}a, it is evident that for $t_H <0.45$, the second-order expansion matches the \emph{true} value closely. However, as $t_H$ increases beyond this range, both the second and third-order expansions begin to deviate from this true value, thereby raising the necessity for the usage of a fourth-order expansion to invoke greater accuracy. Interestingly, for $t_H > 0.45$, the second and the third-order expansions consistently overlap, while for $t_H > 0.75$, the fourth and the fifth-order expansions exhibit similar behavior. In summary, the order of Taylor expansion is determined by analyzing its convergence to the true value, with the second-order expansion being sufficient for small $t_H$, while higher-order expansions are necessary for larger $t_H$ values to maintain the sanctity of the approximation.
\par Furthermore, evaluating Eq. \ref{Taylor2} for any value of $t_H$ (provided $t_V =\pi$) will render in perfectly circular GBZ owing to the absence of skin effect. This may be perceived as a requirement for us to look for other values of $\beta$. To ascertain, what happens when there is skin effect let us examine the validity of our approximations for scenarios where $t_V \neq \pi$ (where $\beta$ depends on $t_H$), say for example, $t_V = 3.5$.
In this case, as mentioned earlier, the skin effect re-emerges, causing 
$\beta$ to vary with $t_H$, which makes it more challenging to assign a specific $\beta$ value due to the involvement of different orders of the expansion. Nevertheless, it is evident that for this value of $t_V$ ($t_V = 3.5$), the wavefunctions get localized toward the left edge of the chain, leading to a GBZ having a radius lesser than unity, that is $0<\beta<1$. Therefore, any value of $\beta$ within this range can serve as a reasonable approximation to the true value. Moreover, any other choice of $\beta$ (within the range $0<\beta<1$), will certainly alter the shape of the functions, yet the value of $t_H$ upto which different orders converge with the true value remains unaltered, highlighting the robustness of our approximation.
\par In Fig. \ref{Figure_12}b, we have plotted the true value versus the ones obtained from different orders of truncation of $E^2(\beta)$ as a function of \(t_H\) for \(t_V = 3.5\), where \(\beta\) has been chosen to lie within the range [0,1]. Interestingly, unlike the \(t_V = \pi\) case with no skin effect, where certain truncation orders (e.g., second and third) overlapped, the presence of the skin effect for \(t_V = 3.5\) leads to significant deviations between different orders of the approximation as \(t_H\) increases. To further validate the approximation, we analyzed the GBZ across different values of $t_H$ using various orders of the expansion. For instance, Fig. \ref{Figure_12}c illustrates the GBZ for $t_H =0.25$. As expected from Fig. \ref{Figure_12}b, every order of expansion coincides well with that of the true value in this regime, indicating that a second-order Taylor expansion is sufficient for obtaining accurate results. On the other hand, Fig. \ref{Figure_12}d presents the GBZ for $t_H = 0.65$. Further, Fig. \ref{Figure_12}b indicates that for $t_H = 0.65$ the second-order expansion suffers significant deviations from the true value, necessitating the utility of employing higher-order expansions. This is corroborated in Fig. \ref{Figure_12}d, where the GBZ obtained from the second-order expansion is notably different from the GBZ derived from the third and fourth-order expansions. Additionally, the GBZs obtained from the fourth and higher-order expansions agree with each other, and with the true value, thus underscoring the requirement for higher-order truncation in this regime.
\par Furthermore, we have computed the coefficients $X_j$ corresponding to $\beta^{j}$ for a third-order Taylor expansion (that is, $n = 3$ in Eq. \ref{Taylor}). The explicit form of the expressions are given as,

\begin{widetext}
    \begin{align}
    X_{(\pm3)} &= \frac{t_H^3}{24} \Big[ \mp \sin(2t_V) \cosh^2(\gamma) + \sinh(2\gamma) [3 \cos^2(t_V) \pm 2]   - \sin^2(t_V) \sinh(2\gamma) + 3 \sin(2t_V) \sinh^2(\gamma)\Big], \\
    X_{(\pm2)} &= \frac{t_H^2}{4} \Big[1 + 2\cos(2t_V) - 2\cosh^2(\gamma) - \cosh(2\gamma) \pm 4\sin(2t_V) \sinh(2\gamma)\Big],
    \end{align}
    \begin{align}
    X_{(\pm1)} &= t_H \Big[ \cosh^2(\gamma) \sin(2t_V) \pm \sinh(2\gamma) \cos^2(t_V) \Big]  \nonumber \\
    & \quad + t_H^3 \Big[ \pm 3 \cosh^2(\gamma) \sin(2t_V) + 2 \sinh(2\gamma) + \cos^2(t_V) \sinh(2\gamma) + 3 \sin^2(t_V) \sinh(2\gamma) \pm \sin(2t_V) \sinh^2(\gamma) \Big], \\
    X_0 &= \cosh^2(\gamma) \sin^2(t_V) - \sinh^2(\gamma) \nonumber \\
    & \quad + 2 t_H^2 \Big[ \cosh^2(\gamma) + \cos^2(t_V) \cosh^2(\gamma) - 2 \cosh^2(\gamma) \sin^2(t_V) + \sinh^2(\gamma) - \sin^2(t_V) \sinh^2(\gamma)\Big].
\end{align}
\end{widetext}
Upon examining these coefficients, it becomes evident that the primary source of deviation between the Taylor-expanded function and its true value is caused by the $X_0$ term, which, nevertheless gets eliminated owing to the minus sign present in Eq. \ref{Taylor2}. As a result, even though the second-order Taylor expansion appears to be valid only for a narrow range of $t_H$ (as shown in Fig. \ref{Figure_12}b) it still provides a good approximation, even with other parameter choices as long as we keep $t_H \le 0.5$.  This is how we have conducted a comprehensive and systematic analysis of each approximation used to develop the Floquet non-Bloch theory.

\end{document}